\documentclass[12pt]{article}
\usepackage{latexsym}
\usepackage{amsmath}
\def\a{\alpha}
\def\b{\beta}

\def\d{\delta}
\def\e{\epsilon}

\def\g{\gamma}
\def\p{\psi}

\def\l{\lambda}
\def\m{\mu}
\def\n{\nu}

\def\s{\sigma}

\def\vf{\varphi}

\def\be{\begin{equation}}
\def\ee{\end{equation}}
\def\arr{\begin{array}{rll}}
\def\ea{\end{array}}
\def\bea{\begin{eqnarray}}
\def\eea{\end{eqnarray}}

\begin{document}
\renewcommand{\thefootnote}{\fnsymbol{footnote}}
\begin{titlepage}
\noindent
\begin{center}
{\Large\bf Quartet unconstrained formulation }\\
\bigskip
{\Large\bf for massless higher spin fields}\\
\bigskip
\vskip 1cm

$
\textrm{\large I.L. Buchbinder\ }^{a} ,~
\textrm{\large A.V. Galajinsky\ }^{b} ,~
\textrm{\large V.A. Krykhtin\ }^{b}
$

\vskip 0.7cm
${}^{a}$ {\it Department of Theoretical Physics, Tomsk State Pedagogical University, \\
634041 Tomsk, Russian Federation}\\
Email: joseph@tspu.edu.ru
\vskip 0.4cm
${}^{b}$ {\it Laboratory of Mathematical Physics, Tomsk Polytechnic University, \\
634050 Tomsk, Russian Federation}\\
{Emails: galajin, krykhtin@mph.phtd.tpu.edu.ru}
\vskip 0.2cm
\end{center}
\vskip 1cm
\begin{abstract}
We construct simple unconstrained
Lagrangian formulations for massless higher spin fields in flat
space of arbitrary dimension and on anti de Sitter background.
Starting from the triplet equations of Francia and
Sagnotti, which describe a chain of spin modes, we introduce an
auxiliary field and find appropriate gauge invariant
constraints that single out the spin--s mode.
The resulting quartet of fields, thus describing an irreducible
representation of the Poincar\'e group, is used to
construct simple Lagrangian formulations, which
are local, free from higher derivative terms and use equal number of
auxiliary fields for an unconstrained description
of any value of spin. Our method
proves to be most efficient
for an unconstrained description of massless higher spin fermions
in anti de Sitter space. A relation of the minimal models with the universal
BRST approach is discussed.

\bigskip
\end{abstract}
\vspace{0.5cm}
PACS: 11.10.z\\
Keywords: higher spin fields
\end{titlepage}
\renewcommand{\thefootnote}{\arabic{footnote}}
\setcounter{footnote}0
\noindent
{\bf 1. Introduction}\\

Recently, the issue of constructing an unconstrained Lagrangian
formulation for massless higher spin fields in flat space and on
(anti) de Sitter background was intensively studied
\cite{pt}--\cite{b3}. By the term "unconstrained" one means a
Lagrangian formulation such that off--shell higher spin gauge
fields and gauge parameters are not subject to any trace
conditions. All the necessary constraints intrinsic to the
conventional approach of Fang and Fronsdal \cite{fr,ff}
arise as equations of motion, possibly after partial gauge fixing,
and follow directly from a Lagrangian.

This line of research is basically motivated by the desire to better
understand a relation between string theory and higher spin gauge
theory. As is well known, in the tensionless limit of string theory
one recovers unconstrained massless higher spin fields (see e.g.
\cite{st}--\cite{lindst}). If string theory is to correspond to a broken phase of
a higher spin gauge theory \cite{v}, an unconstrained description of
massless higher spin fields is desirable. It is also clear that such a
completely unconstrained formulation will be useful for higher spin
field theory itself. Besides, it is believed that an unconstrained
approach may turn out to be useful for studying a possible Lagrangian
formulation for the Vasiliev equations \cite{vas}--\cite{vas2},
which describe interacting massless higher spin fields. We
would like to emphasize here that the construction of a Lagrangian
theory for interacting higher spin fields is one of the principal
unsolved problems of classical field theory (for reviews see e.g.
\cite{r}--\cite{r2}). A more pragmatic reason to search for an
unconstrained formalism  is prompted by possible (perhaps
hypothetical) applications in quantum field theory. It is well known
that Green functions and vertices are calculated by varying an
action with respect to fields. Technically this becomes much
more involved if the latter are constrained.

By now there are two approaches leading to consistent unconstrained
Lagrangian formulations for massless higher spin fields in flat
space and on (anti) de Sitter background. The first approach, known
as the BRST approach, was inspired by string field theory \cite{ost,b},
\cite{pt}--\cite{bkp}.
Here a spin--s field is represented by a
state $|\Phi\rangle$ in an appropriate Fock space and the
conditions, which determine an irreducible massless representation
of the Poincar\'e group, come about as specific operators
annihilating the state. Treating these operators as constrains one
can further construct the canonical BRST charge $Q$. By analogy with
string field theory, an action functional describing a free spin--s
field is built in the form $\int d\mu \langle \Phi |
Q|\Phi\rangle$, where $d\m$ is an appropriate measure (see
\cite{pt}--\cite{bkp} for more details). The cubic
interaction vertices can also be constructed within this approach \cite{b1,bfpt}.
A generalization of the BRST method to the case of massive higher spin fields
was realized in \cite{mass}--\cite{mass2}.

Although the BRST approach automatically yields Lagrangian
formulations in terms of completely unconstrained fields, the
corresponding theories involve a number of auxiliary fields.
Besides, in the case of anti de Sitter background one faces a
complicated problem of constructing a BRST charge associated with a nonlinear algebra of
constraints \footnote{A discussion of this issue can be found
in \cite{nonlin},\cite{nonlin1}.}.
For our
subsequent consideration it is important to stress that the number
of auxiliary fields, i.e. the fields which should be either gauged
away or eliminated with the use of their (algebraic) equations of
motion to reproduce the constrained formulations of Fang and
Fronsdal \cite{fr,ff}, {\it grows} with the value of spin.

Another approach, called the geometric approach, relies upon a
modification of the original Fang--Fronsdal equations
\cite{fs1}--\cite{fms}. Here the trace constraints on the gauge parameter
intrinsic to the conventional formulations of \cite{fr,ff} are forgone
at the price of allowing {\it nonlocal} terms in the equations of
motion. In \cite{fs1} the nonlocal equations were put in an elegant
geometric form by making use of the generalized Riemann curvatures
of de Wit and Freedman \cite{df}. It was also demonstrated that the
nonlocal geometric formulations can be put in a more conventional
local form by introducing an auxiliary field, called the compensator
\cite{fs1}. The compensator plays an important role in this
formalism, as it allows one to avoid the constraint on the gauge
parameter and to readily establish the on--shell equivalence with
the Fang--Fronsdal formulation. Indeed, the gauge transformation law
of the compensator is proportional to a combination, which
specifies the constraint on the gauge parameter within the
Fang--Fronsdal approach. Thus, gauging away the compensator within
the geometric approach one is left with the equations of motion and the
residual gauge symmetry, which are precisely those of the
conventional constrained formulation.

A local version of the geometric
approach incorporates the trace condition on the higher spin gauge
field as the constraint, which enters the action functional with
the corresponding Lagrange multiplier \cite{fs2}. The resulting
formulation, which involves two auxiliary fields for any value of spin,
was called the minimal formulation \cite{fs2}. Notice, however, that the minimal local
Lagrangians of \cite{fs2},\cite{fms} involve {\it higher derivative}
terms\footnote{On--shell the higher derivatives act on the compensator
only and do not alter the Cauchy data for the higher spin gauge field equation of motion.
However, a removal of the higher derivative terms is desirable in order to bring the Lagrangians
to a conventional form.}.

Comparing the two approaches outlined above, it is natural to ask whether
there exists a modified formulation, which is local, free from higher derivative
terms and uses equal number of auxiliary fields for an unconstrained description
of any value of spin. The goal of this paper is to present such a formulation.

To a large extend, the present work was motivated by the previous
study of the tensionless limit of string theory \cite{st} and, in
particular, the triplet equations formulated by Francia and Sagnotti
in \cite{fs,fs2}.
The triplet of fields naturally accommodates
higher spin gauge symmetry with an unconstrained gauge
parameter and describes a chain of irreducible spin modes. It is
important to notice that the bosonic and fermionic triplets allow
simple Lagrangian descriptions \cite{fs}. Notice that in the bosonic case
one of the fields can be eliminated
algebraically without generating higher derivative
terms in the remaining equations of motion \cite{fs,fs2}. This suggests that
the system could be termed equally well a "doublet".
For an earlier consideration of higher spin theories
related to the triplets and doublets see references \cite{b,tet}.

The purpose of this paper is to systematically derive unconstrained
Lagrangian formulations for massless higher spin fields
from the triplets by finding an appropriate
set of gauge invariant constraints, which extract a single
spin--s mode from the chain of irreducible representations.
Interestingly enough, in order to write the
constraints without spoiling the unconstrained gauge symmetry, one
has to introduce the compensator, which is the key ingredient of the
geometric approach. Ultimately, one arrives at a simple Lagrangian
formulation, which is local, free from higher
derivative terms and uses equal number of auxiliary fields for
an unconstrained description of any value of spin. For an integer spin
one has a quartet of fields
and two Lagrange multipliers,
while for a half--integer spin one has a quartet of
fields and three Lagrange
multipliers. Since in the bosonic theory there are six fields one would naturally
call it the "sextet formulation". Analogously, in the fermionic case one would speak about
the "septet formulation". However, since the quartet of fields is shared by both the bosonic and fermionic
models and the Lagrange multipliers prove to vanish on--shell, it seems natural
to stick to the term "quartet formulation" so as to stress the universal feature intrinsic to both the cases.
It should be stressed also that the higher spin theories constructed in this work
can be viewed as an off--shell counterpart of the on--shell reduction \cite{fs1}--\cite{fs2}
from the triplets to the compensator form of higher spin dynamics.
Conventional
Fang--Fronsdal formulations follow after partial gauge fixing and
on--shell elimination of all the auxiliary fields.
Notice that, while the bosonic triplet is well understood both in flat
space and on anti de Sitter background \cite{fs,st}, a consistent formulation of
the fermionic triplet in anti de Sitter space is unknown (see the discussion of this issue
in ref. \cite{st}). Then our method of supplementing the triplets with gauge invariant constraints
provides an efficient means, which allows one to construct a Lagrangian immediately.
The construction of an unconstrained minimal formulation for higher spin fermions in anti de Sitter space
by other methods proves to be much more complicated and was unknown
until quite recently \cite{fms}.
As the BRST approach provides a universal framework for an
unconstrained description of higher spin fields, in this paper
we are also concerned with a precise relation between the
minimal formulations and the BRST models.

The organization of this work is as follows. In sect. 2 we consider massless bosonic higher spin fields in
flat space of arbitrary dimension and put the Fronsdal equation and the gauge transformation in an unconstrained form. Given
a totally symmetric tensor field of rank--s, three auxiliary fields of rank $(s-1)$, $(s-2)$, $(s-3)$ are introduced.
Our unconstrained description relies upon four equations, only two of which are the members of
the bosonic triplet. The third triplet equation comes about as the differential consequence. Thus, the entire system
can be viewed as the bosonic triplet subject to two gauge invariant constraints. A Lagrangian formulation, which
reproduces the unconstrained description of a massless spin--s boson is given in sect. 3. It is argued that
resolving one of the constraints, eliminating two of the auxiliary fields and redefining the
Lagrange multiplier,
one gets the minimal formulation of Francia and Sagnotti \cite{fs2}.
An unconstrained
Lagrangian formulation for a massless spin--s fermion in flat space of arbitrary dimension is constructed in sect. 4.
A totally symmetric tensor field of rank-$(s-\frac{1}{2})$, which now carries an extra Dirac spinor index,
is accompanied by three auxiliary fields of rank-$(s-\frac{3}{2})$, $(s-\frac{5}{2})$, $(s-\frac{5}{2})$,
respectively. An unconstrained version of the Fang--Fronsdal equations relies upon four relations. Only
one of these is the member of the fermionic triplet, while two other triplet equations arise as the
differential consequences. Thus, the entire system can be regarded as the fermionic triplet subject
to three gauge invariant constraints. A link between the minimal unconstrained formulations and
the BRST approach is discussed in sect. 5.
It is shown by explicit calculation that the simple Lagrangian formulations constructed in
this work can be derived from the BRST models by partial gauge fixing and on--shell eliminating some
of the auxiliary fields. In sect. 6
the unconstrained Lagrangian formulations for massless higher spin bosonic and fermionic
fields are generalized to the case of (anti) de Sitter background.
In sect. 7 we summarize our results and discuss possible further developments.

\vspace{0.5cm}

\noindent

{\bf 2. The Fronsdal equation in an unconstrained form}\\

As originally formulated by Fronsdal in \cite{fr}, a free massless spin-s boson is described
by a totally symmetric tensor field $\phi_{\m_1 \dots \m_s}$ subject to the double traceless condition
\be\label{con}
{\phi^{\lambda \s}}_{\lambda \s \m_5 \dots \m_s}=0.
\ee
The equation of motion
\be
\Box \phi_{\m_1 \dots \m_s}-(\partial_{\m_1} \partial^\n \phi_{\n \m_2 \dots \m_s} +\dots)+
(\partial_{\m_1} \partial_{\m_2} {\phi^\n}_{\n \m_3 \dots \m_s}+\dots)=0,
\ee
where the terms implementing the symmetrization of the indices
$\m_1, \dots , \m_s$ are kept implicit, holds invariant under the gauge transformation
\be
\d \phi_{\m_1 \dots \m_s}=\partial_{\m_1} \e_{\m_2 \dots \m_s} + \dots,
\ee
provided the totally symmetric gauge parameter $\e_{\m_1 \dots \m_{s-1}}$ is traceless
\be\label{par}
{\e^\lambda}_{\lambda \m_3 \dots \m_{s-1}}=0.
\ee
Notice that the constraint (\ref{con}) can be easily incorporated into the Fronsdal action by
introducing a Lagrange multiplier. The latter proves to vanish on--shell. However, this does not eliminate
the restriction (\ref{par}) on the gauge
parameter.

For our subsequent consideration it proves convenient to switch to the notation, which suppresses
the vector indices and automatically takes care of symmetrizations. Conventionally, this is done
by introducing an auxiliary variable $y^\m$ such that
\be
\phi_{\m_1 \dots \m_s} (x) \quad \Leftrightarrow \quad \phi^{(s)}(x,y)=\phi_{\m_1 \dots \m_s} (x)~ y^{\m_1} \dots y^{\m_s}.
\ee
Denoting $p_\m=\frac{\partial}{\partial x^\m}$, $\pi_\m=\frac{\partial}{\partial y^\m}$, one
has $\pi^2$ for the trace, $(p\pi)$ for the divergence and $(yp)$ for the derivative of a
field followed by symmetrization of indices. This notations proves to be particularly useful for
establishing gauge invariance of equations of motion or checking their (in)dependence. For example,
the invariance of the Fronsdal equation
\be\label{fron}
\left( p^2-(yp)(p\pi)+(yp)(yp)\frac{\pi^2}{2} \right) \phi^{(s)} (x,y)=0, \quad \pi^2 \pi^2 \phi^{(s)}=0
\ee
under the gauge transformation
\be\label{tra}
\d \phi^{(s)}=(yp) \e^{(s-1)}, \qquad \pi^2 \e^{(s-1)}=0
\ee
is a trivial consequence of the identity
\be
\left( p^2-(yp)(p\pi)+(yp)(yp)\frac{\pi^2}{2} \right) (yp)=(yp)(yp)(yp)\frac{\pi^2}{2}.
\ee
For the reader's convenience we display above the algebra of the Weyl-ordered operators
quadratic in $(p,y,\pi)$.

\begin{table}
\caption{The algebra of the Weyl--ordered operators quadratic in $(p,y,\pi)$}\label{table}
\begin{eqnarray*}
\begin{array}{|l|r|r|r|r|r|c|}
\hline
[ {}{},{}{}] & p^2 & yp & p\pi & \frac{y^2}{2} & \frac{(y\pi+\pi y)}{2} & \frac{\pi^2}{2}
\\
\hline
p^2 & 0 & 0 & 0 & 0 &0 &0\\
\hline
(yp) & 0 & 0 & -p^2 & 0 & -yp & -p\pi\\
\hline
(p\pi) & 0 & p^2 & 0 & yp & p\pi & 0 \\
\hline
\frac{y^2}{2} & 0 & 0 & -yp & 0 & -y^2 & -\frac{(y\pi+\pi y)}{2}\\
\hline
\frac{(y\pi+\pi y)}{2} & 0 & yp & -p\pi & y^2 & 0 & -\pi^2\\
\hline
\frac{\pi^2}{2} & 0 & p\pi & 0 & \frac{(y\pi+\pi y)}{2} & \pi^2 & 0\\
\hline
\end{array}
\end{eqnarray*}
\end{table}
\noindent

Now we are in a position to put the Fronsdal equation (\ref{fron}) and the gauge transformation
(\ref{tra}) in an unconstrained form. To this end, like in \cite{fs1}--\cite{fs2} we
introduce an auxiliary field $\a^{(s-3)}$ whose gauge transformation law is related to the Fronsdal constraint
on the gauge parameter
\be\label{alp}
\d\a^{(s-3)}=\frac{\pi^2}{2} \e^{(s-1)}.
\ee
Then we consider two more auxiliary fields $C^{(s-1)}$, $D^{(s-2)}$ and impose the equations of motion,
which are designed so as to reproduce the Fronsdal formulation, when the compensator $\a^{(s-3)}$ is
gauged away
\bea\label{unc}
&&
p^2 \phi^{(s)} -(yp) C^{(s-1)}=0, \qquad C^{(s-1)}-(p\pi)\phi^{(s)}
+(yp) D^{(s-2)}=0,
\nonumber\\[2pt]
&&
D^{(s-2)}-\frac{\pi^2}{2} \phi^{(s)} +(yp)\a^{(s-3)}=0, \qquad
\frac{\pi^2}{2} D^{(s-2)} -(p\pi)\a^{(s-3)}=0.
\eea
One can easily verify that the system holds invariant under the gauge transformation
\be\label{other}
\d \phi^{(s)}=(yp)\e^{(s-1)}, \quad \d C^{(s-1)} =p^2 \e^{(s-1)},
\quad \d D^{(s-2)}=
(p\pi)\e^{(s-1)},
\ee
with the variation of $\a^{(s-3)}$ as in (\ref{alp}). Notice that the gauge parameter $\e^{(s-1)}$ is
unconstrained.

The equivalence of eqs. (\ref{unc}) to the Fronsdal
formalism is established by imposing the gauge condition
\be
\a^{(s-3)}=0.
\ee
The latter can be implemented with the use of the trace of the gauge parameter. Then the fields
$C^{(s-1)}$ and $D^{(s-2)}$ can be expressed algebraically in terms of the doubly traceless
$\phi^{(s)}$, which brings one back to eq. (\ref{fron}). The residual gauge transformation
\be
\d \a^{(s-3)}=0 \quad \Rightarrow \quad \pi^2 \e^{(s-1)}=0
\ee
is precisely that in eq. (\ref{tra}).

It is obvious from (\ref{unc}) that the fields $C^{(s-1)}$ and $D^{(s-2)}$ are non dynamical and
can be eliminated from the consideration by making use of their equations of motion. What is noteworthy is that
the resulting equations
\bea\label{fs}
&&
\left( p^2-(yp)(p\pi)+(yp)(yp)\frac{\pi^2}{2} \right) \phi^{(s)}=
(yp)(yp)(yp) \a^{(s-3)},
\nonumber\\[2pt]
&&
\frac{\pi^2}{2} \frac{\pi^2}{2} \phi^{(s)}-\left( 2(p\pi)+(yp)\frac{\pi^2}{2} \right) \a^{(s-3)}=0
\eea
are precisely those\footnote{Our $\a^{(s-3)}$ differs from the compensator of Francia and Sagnotti by the
factor $\frac{s(s-1)(s-2)}{2}$.} underlying the local version of the
geometric formulation
by Francia and Sagnotti \cite{fs1}--\cite{fs2}. Although these equations can be derived from an action \cite{fs1,fs2},
they involve higher derivative terms.

Thus, the system (\ref{unc}) can be viewed as a natural extension of the Francia--Sagnotti equations,
which avoids the undesirable higher derivative terms. Notice that in order to cure the higher derivative problem
the use of $D^{(s-2)}$ alone would be sufficient. However, we refrain from eliminating $C^{(s-1)}$ from
the consideration
\footnote{It should be mentioned that the constraints entering eqs. (\ref{unc}) along with other equations
appeared in the context of the BRST consideration of \cite{st}, where they were used to eliminate $C^{(s-1)}$
and $D^{(s-2)}$.}
as it comprises the bosonic triplet, which links
the tensionless limit of the BRST--quantized bosonic string theory and the massless higher spin gauge
theory \cite{fs,st,b} (for an earlier discussion of higher spin theories
related to the triplets see \cite{b,tet}).
Worth mentioning also that in implicit form the set of fields entering
(\ref{unc}) appeared in \cite{g}. However, a Lagrangian formulation for (\ref{unc}) was not discussed.
In the next section we explicitly construct a Lagrangian density,
which reproduces equations (\ref{unc}) and provides a simple unconstrained description of a massless spin--s boson.

\vspace{0.5cm}
\noindent

{\bf 3. Simple unconstrained Lagrangian for a massless spin-s boson}\\

We begin this section by discussing the differential consequence of eqs. (\ref{unc})
\be\label{supp}
p^2 D^{(s-2)}-(p\pi) C^{(s-1)}=0.
\ee
It turns out that a local unconstrained Lagrangian formulation for spin--s boson, which is
free from higher derivative terms incorporates also the equation (\ref{supp}).
Interestingly enough, omitting the second line in (\ref{unc}), i.e. the trace conditions, and
imposing the equation (\ref{supp}) one gets precisely the bosonic triplet of \cite{fs,st}. As is well known,
the triplet describes a chain of spins $s,s-2, \dots, 0$ or $1$ depending on whether $s$
is even or odd. Thus, the restrictions that comprise the second line in (\ref{unc}) can be viewed as
gauge invariant constraints that single out the spin--s mode from the chain of fields
described by the triplet. Notice that the first of the constraints appeared previously in \cite{fs,st}.

The discussion above suggests that in order to build an unconstrained Lagrangian formulation
for a massless
spin-s boson it suffices to start with the Lagrangian density corresponding to the triplet (see e.g. \cite{st})
and enforce the second line in (\ref{unc}) by introducing two Lagrange multipliers $\lambda^{(s-2)}$,
$\lambda^{(s-4)}$. In component form the action reads
\bea\label{action}
&&
S=-{(-1)}^s \int d^N x \Bigl\{ \textstyle{\frac{1}{2}} \phi_{\m_1 \dots \m_s} \Box {}  \phi^{\m_1 \dots \m_s}-
s\phi_{\m_1 \dots \m_s}  \partial^{\m_1} C^{\m_2 \dots \m_s} -
\textstyle{\frac{1}{2}} s C_{\m_1 \dots \m_{s-1}} C^{\m_1 \dots \m_{s-1}} -
\nonumber\\[2pt]
&&
\quad \qquad -s(s-1)C_{\m_1 \dots \m_{s-1}}\partial^{\m_1} D^{\m_2 \dots \m_{s-1}} -
\textstyle{\frac{1}{2}} s(s-1) D_{\m_1 \dots \m_{s-2}} \Box D^{\m_1 \dots \m_{s-2}}+
\nonumber\\[2pt]
&&
\quad \qquad +\lambda^{\m_1 \dots \m_{s-2}} \left( \textstyle{\frac{1}{2}} {\phi^{\n}}_{\n \m_1 \dots \m_{s-2}}
-D_{\m_1 \dots \m_{s-2}} -\textstyle{\frac{1}{2}} (s-2)\partial_{\m_1} \a_{\m_2 \dots \m_{s-2}}\right)+
\nonumber\\[2pt]
&&
\quad \qquad +
\lambda^{\m_1 \dots \m_{s-4}} \left({D^\n}_{\n \m_1 \dots \m_{s-4}}-\partial^\n \a_{\n \m_1 \dots \m_{s-4}}\right)
\Bigr\}
\eea
and the corresponding gauge transformation is of the form
\bea\label{GT}
&&
\d \phi_{\m_1 \dots \m_s}=\partial_{\m_1} \e_{\m_2 \dots \m_s}+\dots,
\qquad \d C_{\m_1 \dots \m_{s-1}}=\Box \e_{\m_1 \dots \m_{s-1}},
\nonumber\\[2pt]
&&
\d D_{\m_1 \dots \m_{s-2}}=\partial^\n
\e_{\n \m_1 \dots \m_{s-2}}, \qquad \d \a_{\m_1 \dots \m_{s-3}}={\e^{\n}}_{\n \m_1 \dots \m_{s-3}}.
\eea

Now one has to check that the introduction of the Lagrange multipliers does not alter
eqs. (\ref{unc}) and (\ref{supp}). Variation of the action (\ref{action}) with respect to all the
unconstrained fields and Lagrange multipliers gives the equations of motion
\bea\label{0}
&& p^2 \phi^{(s)}-(yp) C^{(s-1)}+\frac{y^2}{2} \lambda^{(s-2)}=0,
\\[2pt]\label{1}
&&
C^{(s-1)}-(p\pi)\phi^{(s)} +(yp) D^{(s-2)}=0,
\\[2pt]\label{2}
&&
D^{(s-2)}-\frac{\pi^2}{2} \phi^{(s)} +(yp)\a^{(s-3)}=0,
\\[2pt]\label{3}
&&
\frac{\pi^2}{2} D^{(s-2)} -(p\pi)\a^{(s-3)}=0,
\\[2pt]\label{4}
&&
p^2 D^{(s-2)}-(p\pi) C^{(s-1)}+\lambda^{(s-2)}-y^2 \lambda^{(s-4)}=0,
\\[2pt]\label{6}
&&
\frac{1}{2} (p\pi)\lambda^{(s-2)}+(yp) \lambda^{(s-4)}=0,
\eea
where the field redefinition
\be
s C^{(s-1)} \rightarrow C^{(s-1)}, \qquad s(s-1) D^{(s-2)} \rightarrow D^{(s-2)},
\qquad \frac{s(s-1)(s-2)}{2} \a^{(s-3)} \rightarrow \a^{(s-3)}
\ee
was implemented so as to fit the notation of the previous section.

In order to facilitate the subsequent analysis of the Lagrange multipliers, let us
first discuss a technical issue. Consider the equation
\be\label{l3}
\frac{\pi^2}{2}\frac{y^2}{2} \psi^{(s)}=0,
\ee
where $\psi^{(s)}$ is an arbitrary rank-s tensor. On account of the
identity
\be\label{id}
\frac{\pi^2}{2}\frac{y^2}{2} \p^{(s)}=\left(s+\frac{N}{2}\right)\p^{(s)}+\frac{y^2}{2}\frac{\pi^2}{2} \p^{(s)},
\ee
where $N$ stands for the dimension of space--time, one concludes that
$\psi^{(s)}$ is proportional to its trace. Successive multiplications of (\ref{id})
by $\frac{\pi^2}{2}$ relate the trace of $\psi^{(s)}$ to its double trace, the double trace of $\psi^{(s)}$
to its triple trace etc. Clearly, this process terminates at a final step. Going backward one finds
\be\label{ll3}
\psi^{(s)}=0.
\ee

In order to demonstrate that $\lambda^{(s-2)}$ and $\lambda^{(s-4)}$ vanish on--shell, it suffices to
examine two differential
consequences of (\ref{0})--(\ref{4})~\footnote{When obtaining (\ref{l2}), it proves convenient to
make use of the relation $p^2 \a^{(s-3)}-\frac{\pi^2}{2} C^{(s-1)}=0$, which follows from (\ref{1}), (\ref{2})
and (\ref{3}).}
\be\label{l2}
\lambda^{(s-2)}-y^2 \lambda^{(s-4)}=\frac{\pi^2}{2} \frac{y^2}{2}\lambda^{(s-2)},
\quad
\frac{\pi^2}{2} \left( \lambda^{(s-2)} -   y^2 \lambda^{(s-4)}\right)=0.
\ee
From (\ref{l2}) one immediately obtains the restriction
\be
\frac{\pi^2}{2} \frac{\pi^2}{2} \frac{y^2}{2}\lambda^{(s-2)}=0,
\ee
which means that $\lambda^{(s-2)}$ is traceless
\be\label{tr}
\frac{\pi^2}{2} \lambda^{(s-2)}=0.
\ee
This is proved by applying the same kind of reasoning as above, when passing from
(\ref{l3}) to (\ref{ll3}).
Being combined with the second equation in (\ref{l2}), this gives
\be
\lambda^{(s-4)}=0,
\ee
which, in view of the first equation in (\ref{l2}) and (\ref{tr}), causes $\lambda^{(s-2)}$ to vanish
\be
\lambda^{(s-2)}=0.
\ee

As was mentioned above, eliminating the fields $C^{(s-1)}$, $D^{(s-2)}$ from the equations of motion (\ref{unc})
one obtains the minimal unconstrained formulation of \cite{fs2}.
Notice that the corresponding removal of the fields
from the action functional (\ref{action}) does not yield directly the action in \cite{fs2}.
The reason is that in  \cite{fs2} the
spin--$(s-4)$ Lagrange multiplier transforms under the gauge transformations, while in our formulation
it holds invariant. Thus, the two models are related by a field redefinition,
which makes the Lagrange multiplier in \cite{fs2} gauge invariant (see eq. (3.12) in \cite{fs2}).

To summarize, the action functional (\ref{action}) yields equations (\ref{unc})
of the previous section and, as thus, provides a simple unconstrained Lagrangian formulation for a
massless spin-s boson. It should be appreciated that, in contrast to other approaches,
this formulation is local, free from higher derivative terms and uses equal number of
auxiliary fields for any value of spin.

\vspace{0.5cm}
\noindent

{\bf 4. Simple unconstrained Lagrangian for a massless spin--s fermion}\\

According to Fang and Fronsdal \cite{ff}, a massless spin--s fermion is described by a totally symmetric
rank-$(s-\frac{1}{2})$ tensor field, which carries an extra Dirac spinor index $A$
\be
\Psi_{A \m_1 \dots \m_{s-\frac{1}{2}}} (x) \quad
\Leftrightarrow \quad \Psi^{(s-\frac{1}{2})}_A (x,y)=\Psi_{A \m_1 \dots \m_{s-\frac{1}{2}}} (x) ~ y^{\m_1}
\dots {y}^{\m_{s-\frac{1}{2}}}.
\ee
In what follows we keep the spinor indices implicit and denote $s-\frac{1}{2}=n$. The equation of motion
\be\label{fang}
\left( (\g p)-(yp) (\g\pi) \right) \Psi^{(n)}=0,
\ee
where $\g^\m$ stand for the Dirac matrices~\footnote{The Dirac matrices
satisfy the standard identity $\g_\m \g_\n+\g_\n \g_\m=2\eta_{\m\n}$, where $\eta_{\m\n}$ is the Minkowski metric
in an $N$--dimensional space--time with the signature $(+,-,\dots,-)$. We choose a representation of
the $\g$--matrices such that
${\g_0}^{+}=\g_0$, ${(\g_0\g_\m)}^{+}=\g_0\g_\m$. Then for arbitrary spinors $\psi$ and $\chi$ one has
${(\bar\psi\chi)}^{+}=\bar\chi\psi$,
${(\bar\psi\g_\m \chi)}^{+}=\bar\chi \g_\m\psi$.}, and the constraint
\be\label{cons1}
(\g\pi)(\g\pi)(\g\pi)\Psi^{(n)}=0
\ee
hold invariant under the gauge transformation
\be\label{vp}
\d \Psi^{(n)}=(yp) \e^{(n-1)},
\ee
provided the gauge parameter $\e^{(n-1)}$ is $\g$--traceless
\be\label{cons2}
(\g\pi) \e^{(n-1)}=0.
\ee
Our goal is to relax the constraints (\ref{cons1}), (\ref{cons2}) by introducing auxiliary fields and to
construct an appropriate Lagrangian density.

By analogy with the bosonic case one starts with a compensator
$\a^{(n-2)}$ whose transformation law is related to the constraint (\ref{cons2})
\be\label{va}
\d \a^{(n-2)}=(\g\pi) \e^{(n-1)}.
\ee
Then one introduces two auxiliary spin--tensors $C^{(n-1)}$, $D^{(n-2)}$
and imposes the equations of motion, which are designed so as to reproduce the Fang--Fronsdal equations
(\ref{fang}), (\ref{cons1}), when the compensator $\a^{(n-2)}$ is gauged away
\bea\label{uncf}
&&
(\g p)\Psi^{(n)} -(yp) C^{(n-1)}=0, \quad
C^{(n-1)}-(\g\pi) \Psi^{(n)} +(yp)\a^{(n-2)}=0,
\nonumber\\[2pt]
&&
D^{(n-2)}-\frac{1}{2} (\g\pi)C^{(n-1)}-\frac{1}{2}(\g p) \a^{(n-2)}=0, \quad
(\g\pi) D^{(n-2)}-(p\pi)\a^{(n-2)}=0.
\eea
Accompanying the transformation laws (\ref{vp}) and (\ref{va}) by
\be
\d C^{(n-1)}=(\g p) \e^{(n-1)},
\quad \d D^{(n-2)}=(p \pi) \e^{(n-1)},
\ee
with an unconstrained gauge parameter $\e^{(n-1)}$,
one can demonstrate that the system (\ref{uncf}) is invariant.

In order to construct a Lagrangian formulation reproducing equations (\ref{uncf}), we first consider
their differential consequences
\be\label{fsa}
(\g p) C^{(n-1)} -(p\pi) \Psi^{(n)} + (yp) D^{(n-2)}=0, \quad
(\g p) D^{(n-2)}-(p\pi) C^{(n-1)}=0.
\ee
An important observation is that, being combined with the first equation in (\ref{uncf}), these
give precisely the fermionic triplet considered in \cite{fs} (see also \cite{st} for
a related discussion of the triplet in the context of string theory).
As the triplet describes a chain of half--integer spins \cite{fs}, the last three equations in
(\ref{uncf}) can be viewed as gauge invariant constraints extracting the spin--s mode.

The discussion above also makes clear how to construct an unconstrained
Lagrangian formulation for a massless spin--s fermion. One has to take
the Lagrangian density describing the triplet (see e.g. \cite{fs}) and enforce the last three equations in
(\ref{uncf}) by introducing the Lagrange multipliers $\lambda^{(n-1)}$,
$\lambda^{(n-2)}$, $\lambda^{(n-3)}$
\bea\label{actionf}
&&
S=\int d^N x \Bigl\{
i\bar\Psi_{\m_1 \dots\m_n} \left( \slash\!\!\!\partial~ \Psi^{\m_1 \dots\m_n}
-n\partial^{\m_1} C^{\m_2 \dots \m_n}+n\g^{\m_1} \lambda^{\m_2 \dots \m_n}
\right)+in{\bar C}_{\m_1 \dots\m_{n-1}}\left( \slash\!\!\!\partial~ C^{\m_1 \dots\m_{n-1}}-
\right.
\nonumber\\[2pt]
&&
\left.
\quad
-\partial^\n {\Psi_\n}^{\m_1 \dots\m_{n-1}}
+(n-1)\partial^{\m_1} D^{\m_2 \dots \m_{n-1}}-\lambda^{\m_1 \dots\m_{n-1}}
+\textstyle{\frac{1}{2}}(n-1)\g^{\m_1} \lambda^{\m_2 \dots \m_{n-1}}\right)+
\nonumber\\[2pt]
&& \quad
+in(n-1){\bar D}_{\m_1 \dots\m_{n-2}}\left(-\slash\!\!\!\partial~ D^{\m_1 \dots\m_{n-2}}+\partial^\n
{C_\n}^{\m_1 \dots\m_{n-2}}-\lambda^{\m_1 \dots\m_{n-2}}-(n-2)\g^{\m_1} \lambda^{\m_2 \dots \m_{n-2}}\right)+
\nonumber\\[2pt]
&& \quad
+in(n-1){\bar\a}_{\m_1 \dots \m_{n-2}} \left(-\textstyle{\frac{1}{2}} \slash\!\!\!\partial~ \lambda^{\m_1 \dots \m_{n-2}}
+\partial^\n {\lambda_\n}^{\m_1 \dots \m_{n-2}}-(n-2)\partial^{\m_1} \lambda^{\m_2 \dots\m_{n-2}}
\right)+
\nonumber\\[2pt]
&&
\quad +in {\bar \lambda}_{\m_1 \dots\m_{n-1}} \left(C^{\m_1 \dots\m_{n-1}}-\gamma^\n {\Psi_\n}^{\m_1 \dots \m_{n-1}}
+(n-1) \partial^{\m_1} \a^{\m_2 \dots \m_{n-1}} \right)+
\nonumber\\[2pt]
&&
\quad
+in(n-1){\bar\lambda}_{\m_1 \dots \m_{n-2}} \left(D^{\m_1 \dots \m_{n-2}}-
\textstyle{\frac{1}{2}} \gamma^\n {C_\n}^{\m_1 \dots \m_{n-2}}
-\textstyle{\frac{1}{2}} \slash\!\!\!\partial~ \a^{\m_1 \dots \m_{n-2}} \right)+
\nonumber\\[2pt]
&&
\quad +in(n-1)(n-2) {\bar \lambda}_{\m_1 \dots\m_{n-3}} \left( \gamma^\n {D_\n}^{\m_1 \dots \m_{n-3}}-
\partial^\n {\a_\n}^{\m_1 \dots \m_{n-3}}\right) \Bigr\}.
\eea
This action is invariant under the gauge transformation
\bea
&&
\d \Psi_{\m_1 \dots \m_n}=\partial_{\m_1} \e_{\m_2 \dots \m_n}+\dots, \qquad
\d C_{\m_1 \dots \m_{n-1}}=\slash\!\!\!\partial~\e_{\m_1 \dots \m_{n-1}},
\nonumber\\[2pt]
&&
\d D_{\m_1 \dots \m_{n-2}}=\partial^\n \e_{\n\m_1 \dots \m_{n-2}}, \qquad \d \a_{\m_1 \dots \m_{n-2}}=\g^\n
\e_{\n \m_1 \dots \m_{n-2}}.
\eea
The corresponding equations of motion read\footnote{In order to simplify the equations, we redefine the fields as follows
$n C^{(n-1)} \rightarrow  C^{(n-1)}$, $ n(n-1) D^{(n-2)} \rightarrow  D^{(n-2)}$,
$n(n-1) \a^{(n-2)} \rightarrow  \a^{(n-2)}$, $n \lambda^{(n-1)} \rightarrow  \lambda^{(n-1)}$,
$n(n-1) \lambda^{(n-2)} \rightarrow  \lambda^{(n-2)}$, $n(n-1)(n-2) \lambda^{(n-3)} \rightarrow
\lambda^{(n-3)}$.}
\bea\label{f0}
&&
(\g p)\Psi^{(n)}-(yp)C^{(n-1)}+(\g y)\lambda^{(n-1)}=0,
\\[2pt]\label{f1}
&&
(\g p)C^{(n-1)}-(p\pi)\Psi^{(n)}+(yp) D^{(n-2)}-\lambda^{(n-1)}+\frac{1}{2}(\g y) \lambda^{(n-2)}=0,
\\[2pt]\label{f2}
&&
(\g p)D^{(n-2)}-(p\pi)C^{(n-1)}+\lambda^{(n-2)}+(\g y) \lambda^{(n-3)}=0,
\eea
\bea\label{f3}
&&
C^{(n-1)}-(\g\pi)\Psi^{(n)}+(yp)\a^{(n-2)}=0,
\\[2pt]\label{f4}
&&
\frac{1}{2}(\g p) \a^{(n-2)}+\frac{1}{2}(\g\pi)C^{(n-1)}-D^{(n-2)}=0,
\\[2pt]\label{f5}
&&
(\g\pi)D^{(n-2)}-(p\pi)\a^{(n-2)}=0,
\\[2pt]\label{f6}
&&
\frac{1}{2}(\g p)\lambda^{(n-2)}-(p\pi)\lambda^{(n-1)}+(yp)\lambda^{(n-3)}=0.
\eea

In order to demonstrate that the Lagrange multipliers $\lambda^{(n-1)}$,
$\lambda^{(n-2)}$, $\lambda^{(n-3)}$ vanish on--shell, we act by the operator $(\g\pi)$ on (\ref{f0})--(\ref{f2})
and then make proper use of (\ref{f0})--(\ref{f5}) to get the restrictions
\bea\label{la1}
&&
\lambda^{(n-1)}-\frac{1}{2}(\g y)\lambda^{(n-2)}=\frac{1}{2}(\g\pi) (\g y)\lambda^{(n-1)},
\\[2pt]\label{la2}
&&
\lambda^{(n-2)}+(\g y)\lambda^{(n-3)}=(\g\pi)\left(
\lambda^{(n-1)}-\frac{1}{2}(\g y)\lambda^{(n-2)}\right),
\\[2pt]\label{la3}
&&
(\g\pi)\left(\lambda^{(n-2)}+(\g y)\lambda^{(n-3)}\right)=0.
\eea
Acting by the operators $(\g\pi)(\g\pi)$ and $(\g\pi)$ on (\ref{la1}) and (\ref{la2}), respectively,
and taking into account (\ref{la3}), one finds
\be
(\g\pi)(\g\pi)(\g\pi)(\g y) \lambda^{(n-1)}=0.
\ee
From here it follows that $\lambda^{(n-1)}$ is double $\g$--traceless
\be\label{n-1}
(\g\pi)(\g\pi)\lambda^{(n-1)}=\pi^2 \lambda^{(n-1)}=0.
\ee
The proof is similar to the bosonic case (see eqs. (\ref{l3})--(\ref{ll3}))
and relies upon
the identity
\be
(\g\pi)(\g y) \varphi^{(n)}=\left(2n +N\right) \varphi^{(n)}-(\g y)(\g\pi) \varphi^{(n)},
\ee
which is valid for an arbitrary rank--n spin-tensor $\varphi^{(n)}$. Comparing the latter with
(\ref{id}), we see that $(\g y)$ and $(\g\pi)$ are the fermionic analogues of $\frac{y^2}{2}$ and
$\frac{\pi^2}{2}$, respectively.

Combining equations (\ref{la2}), (\ref{la3}) with (\ref{n-1}) one then gets the chain of relations
\be\label{n-2}
(\g\pi)(\g\pi)(\g y) \lambda^{(n-2)}=0 ~ \rightarrow ~(\g\pi)\lambda^{(n-2)}=0 ~\rightarrow ~
(\g\pi)(\g y) \lambda^{(n-3)}=0 ~ \rightarrow ~\lambda^{(n-3)}=0.
\ee
At this point eq. (\ref{la2}) can be used to relate $\lambda^{(n-2)}$ with the $\g$--trace of $\lambda^{(n-1)}$
\be\label{n-3}
\left( 2(n-1)+N \right) \lambda^{(n-2)}=2(\g\pi)\lambda^{(n-1)}.
\ee
Notice that the last line is fully consistent with (\ref{n-1}) and (\ref{n-2}).
Finally, substituting
(\ref{n-3}) in (\ref{la1}) one can fix $\lambda^{(n-1)}$ and $\lambda^{(n-2)}$
\be
(\g\pi)(\g y) \lambda^{(n-1)}=0 ~ \rightarrow ~\lambda^{(n-1)}=0 ~ \rightarrow ~\lambda^{(n-2)}=0.
\ee

Thus, all the Lagrange multipliers vanish on--shell and the action functional (\ref{actionf})
correctly reproduces the unconstrained description (\ref{uncf}) of a massless fermion field with
spin $s=n+\frac{1}{2}$.

\vspace{0.5cm}
\noindent

{\bf 5. A relation with the BRST approach}\\

As was mentioned in the Introduction, the BRST approach provides a universal method of constructing
unconstrained Lagrangian formulations for higher spin fields. It is then interesting to see how
the models considered in the previous sections are related with those built within
the BRST approach. Below we demonstrate that the former follow from the latter
after partial gauge fixing and eliminating some of auxiliary fields with the use of
their equations of motion. For simplicity we discuss in detail the integer spin case.
Half--integer spins can be treated in a similar fashion. In contrast to the previous
discussion of the triplets in the context of the BRST approach \cite{st},
we obtain the minimal formulation directly at the Lagrangian level. Besides,
our method of gauge fixing is simpler than in \cite{st}.

Within the framework of the BRST approach
a spin--s field is represented by a state in the Fock
space\footnote{$a^{+}_\m$, $a_\m$ are the standard creation and annihilation operators
obeying the commutation relation $\bigl[\,a_\mu,a_\nu^+\bigr]=-\eta_{\mu\nu}$, with
$\eta_{\mu\nu}=\mbox{diag}(+,-,\ldots,-)$.}
\begin{eqnarray}
|\phi\rangle
&=&
\frac{(-i)^s}{s!}\;\phi(x)_{\mu_1\ldots\mu_s}
a^{\mu_1+}\ldots a^{\mu_s+}|0\rangle,
\qquad
-a_\nu^+a^\nu\;|\phi\rangle=s\;|\phi\rangle.
\label{phi}
\end{eqnarray}
In order to describe irreducible representations of the Poincar\'e group, one introduces
the operators
\be\label{oper}
l_0=-p^2, \quad l_1=p^\mu{}a_\mu, \quad l_1^+=p^\mu{}a_\mu^+, \quad l_2=\frac{1}{2}a_\mu{}a^\mu,
\quad l_2^+=\frac{1}{2}a_\mu^+{}a^{\mu+},
\ee
where $p_\mu=-i\frac{\partial}{\partial x^\mu}$ , and imposes the restrictions
\begin{eqnarray}
l_0|\phi\rangle
=l_1|\phi\rangle
=l_2|\phi\rangle
=0.
\label{Eqs}
\end{eqnarray}
Regarding these restrictions as constraints one can construct the nilpotent BRST charge
(see e.g. \cite{bpt})
\begin{eqnarray}
Q
&=&
\eta_0l_0+\eta_1^+l_1+\eta_1l_1^+
+\eta_2^+L_2+\eta_2L_2^+
-\eta_1^+\eta_1{\cal{}P}_0
+\eta_2^+\eta_1{\cal{}P}_1
+\eta_1^+\eta_2{\cal{}P}_1^+
,
\label{Q}
\end{eqnarray}
where, along with the ghost sector the original Fock space was extended by an auxiliary pair of
creation and annihilation
operators $b^+$, $b$ (with $[b,b^+]=1$). In (\ref{Q}) we made use of the notation
\begin{eqnarray}
&&
L_2=l_2+b^+b^2-b\sigma,
\qquad
L_2^+=l_2^++b^+
,
\\
&&
\sigma
=
-a_\nu^+a^\nu+\frac{N}{2}+2b^+b
+\eta_1^+{\cal{}P}_1-\eta_1{\cal{}P}_1^+
+2\eta_2^+{\cal{}P}_2-2\eta_2{\cal{}P}_2^+
\label{sigma}
\quad \Rightarrow
\quad
[Q,\sigma]=0
.
\end{eqnarray}
The ghost operators are supposed to obey the anti commutation relations
\begin{eqnarray}
&&
\{\eta_0,{\cal{}P}_0\}=
\{\eta_1,{\cal{}P}_1^+\}=
\{\eta_1^+,{\cal{}P}_1\}=
\{\eta_2,{\cal{}P}_2^+\}=
\{\eta_2^+,{\cal{}P}_2\}=1
\label{comgh}
\end{eqnarray}
and the vacuum state in the ghost sector is specified by the relations
\begin{eqnarray}
&&
\mathcal{P}_0|0\rangle
=\eta_1|0\rangle
=\mathcal{P}_1|0\rangle
=\eta_2|0\rangle
=\mathcal{P}_2|0\rangle
=0,
\nonumber\\[2pt]
&&
\eta_0|0\rangle\neq0,
\quad
\eta_1^+|0\rangle\neq0,
\quad
\mathcal{P}_1^+|0\rangle\neq0,
\quad
\eta_2^+|0\rangle\neq0,
\quad
\mathcal{P}_2^+|0\rangle\neq0
.
\label{ghvac}
\end{eqnarray}
The BRST operator (\ref{Q}) acts
on states depending on both $a^{+\mu}$, $b^+$ and the
ghost operators $\eta_0$,
$\eta_1^+$, ${\cal{}P}_1^+$, $\eta_2^+$, ${\cal{}P}_2^+$.
Such a generic state reads
\begin{eqnarray}
|\Psi\rangle
&=&
\sum_{k_i}
(b^+)^{k_2}
(\eta_0)^{k_3}
(\eta_1^+)^{k_4} ({\cal{}P}_1^+)^{k_5}
(\eta_2^+)^{k_6} ({\cal{}P}_2^+)^{k_7}
a^{+\mu_1}\cdots a^{+\mu_{k_1}}
\Phi_{\mu_1\cdots\mu_{k_1}}^{k_2\cdots{}k_7}(x)|0\rangle,
\label{Psi}
\end{eqnarray}
with $k_1$, $k_2$ running from 0 to $\infty$ and $k_3$, $k_4$, $k_5$,
$k_6$, $k_7$ taking the values 0 or 1.

The irreducibility conditions (\ref{Eqs}) are encoded in
the equations
\begin{align}
&
Q|\Psi\rangle=0
,
&&
gh(|\Psi\rangle)=0,
\label{QPsi}
\\
&
\sigma|\Psi\rangle=(s+\frac{N-6}{2})|\Psi\rangle
\label{sPsi},
\end{align}
which exhibit the reducible gauge symmetry
\begin{align}
\label{dPsi}
&
\delta|\Psi\rangle=Q|\Lambda\rangle,
&&
&
\sigma|\Lambda\rangle=(s+\frac{N-6}{2})|\Lambda\rangle,
&&
gh(|\Lambda\rangle)=-1,
\\
\label{dLambda}
&
\delta|\Lambda\rangle=Q|\Omega\rangle,
&&
&
\sigma|\Omega\rangle=(s+\frac{N-6}{2})|\Omega\rangle,
&&
gh(|\Omega\rangle)=-2.
\end{align}
Notice that (\ref{sPsi}) is imposed so as
$|\Psi\rangle$ to describe a single spin-s mode.
This is an extension of the second equation in (\ref{phi})
to the enlarged Fock space. Had we discarded ~(\ref{sPsi}),
$|\Psi\rangle$ would have been the infinite sum of
fields with the increasing value of integer spin.
As $[Q,\sigma]=0$, the equations (\ref{QPsi}) and
(\ref{sPsi}) are compatible.

Finally, with the use of the BRST charge one can construct the Lagrangian
\begin{eqnarray}
-\mathcal{L}=\int\!\! d\eta_0\; \langle\Psi|KQ|\Psi\rangle
\label{Qaction}
\end{eqnarray}
which describes a massless spin--s boson.
The Fronsdal formulation follows from (\ref{Qaction}) after partial gauge fixing and
eliminating all the auxiliary fields.
In (\ref{Qaction}) the standard scalar
product in the Fock space is used and $K$ is a specific invertible operator providing the
reality of the Lagrangian (see e.g. \cite{bpt} for more details). The latter
acts as the unit operator in the entire Fock space, but for the sector
controlled by ($b^+$,~$b$).

Now let us show that after partial gauge fixing and on--shell elimination of some of the
auxiliary fields one can derive the action functional
(\ref{action}) from (\ref{Qaction}).
The idea is first to fix the reducible gauge symmetry of the formalism and then
gauge away the $b^+$--dependent parts of fields where it proves possible.
To this end, let us represent the
states and the gauge parameters as power series in the ghost operators
\begin{eqnarray}
|\Psi\rangle&=&
|S_1\rangle
+\eta_1^+{\cal{}P}_1^+|S_2\rangle
+\eta_1^+{\cal{}P}_2^+|S_3\rangle
+\eta_2^+{\cal{}P}_1^+|S_4\rangle
+\eta_2^+{\cal{}P}_2^+|S_5\rangle
+\eta_1^+\eta_2^+{\cal{}P}_1^+{\cal{}P}_2^+|S_6\rangle
\nonumber
\\
&&{}
+\eta_0{\cal{}P}_1^+|A_1\rangle
+\eta_0{\cal{}P}_2^+|A_2\rangle
+\eta_0\eta_1^+{\cal{}P}_1^+{\cal{}P}_2^+|A_3\rangle
+\eta_0\eta_2^+{\cal{}P}_1^+{\cal{}P}_2^+|A_4\rangle
,
\label{Psi-}
\\
|\Lambda\rangle&=&
{\cal{}P}_1^+|\lambda_1\rangle
+{\cal{}P}_2^+|\lambda_2\rangle
+\eta_1^+{\cal{}P}_1^+{\cal{}P}_2^+|\lambda_3\rangle
+\eta_2^+{\cal{}P}_1^+{\cal{}P}_2^+|\lambda_4\rangle
+\eta_0{\cal{}P}_1^+{\cal{}P}_2^+|\lambda_5\rangle,
\\
|\Omega\rangle&=&
{\cal{}P}_1^+{\cal{}P}_2^+|\omega\rangle,
\end{eqnarray}
where the vectors $|S_i\rangle$, $|A_i\rangle$,
$|\lambda_i\rangle$, $|\omega\rangle$ depend on $a^{\mu+}$ and
$b^+$ only.

In order to get rid of the reducibility of the theory,
one uses the gauge transformation law of the parameter
$|\lambda_1\rangle$
\begin{eqnarray}
\delta|\lambda_1\rangle=-(l_2^++b^+)|\omega\rangle.
\end{eqnarray}
From here it follows that $|\omega\rangle$ can be used to gauge away
the $b^+$--dependent part of $|\lambda_1\rangle$. As $|\omega\rangle$
is now fixed, the residual gauge transformation is irreducible, with
$|\lambda_2\rangle$,
$|\lambda_3\rangle$, $|\lambda_4\rangle$, $|\lambda_5\rangle$ being unconstrained and
$|\lambda_1\rangle$ obeying the restriction
$b|\lambda_1\rangle=0$. The latter condition can also be written as
\be
\mathcal{P}_2^+b|\Lambda\rangle=0.
\ee

Now we turn to gauge transformation laws of the fields and
gauge away the $b^+$--dependent parts where it proves possible.
For
$|S_1\rangle$, $|S_2\rangle$, $|S_4\rangle$, $|A_1\rangle$ one readily obtains
the relations
\bea\label{dS1}
&&
\delta|S_1\rangle=
  (l_2^++b^+)|\lambda_2\rangle
  +l_1^+|\lambda_1\rangle, \qquad
\delta|S_2\rangle=
  (l_2^++b^+)|\lambda_3\rangle
  +l_1|\lambda_1\rangle
  -|\lambda_2\rangle,
\nonumber\\[2pt]
&&
\delta|S_4\rangle=
   (l_2^++b^+)|\lambda_4\rangle
   +l_2|\lambda_1\rangle, \qquad
\delta|A_1\rangle=
  (l_2^++b^+)|\lambda_5\rangle
  +l_0|\lambda_1\rangle.
\eea
From here we see that making use of the gauge parameters
$|\lambda_2\rangle$,
$|\lambda_3\rangle$, $|\lambda_4\rangle$, $|\lambda_5\rangle$ one can
eliminate the $b^+$--dependent parts of the states $|S_1\rangle$, $|S_2\rangle$,
$|S_4\rangle$, $|A_1\rangle$, respectively. This gauge choice can be conveniently written in the form
\be
\mathcal{P}_2^+b|\Psi\rangle=0.
\ee
At this stage one is left with the single gauge parameter $|\lambda_1\rangle$ subject to the restriction
$b|\lambda_1\rangle=0$.

A relation with the unconstrained Lagrangian formulation for a massless spin--s boson
constructed in sect. 3 is established via the identification of the fields and the gauge parameters
\begin{align}
&
|S_1\rangle=|\phi\rangle,
&&
|S_2\rangle=|D\rangle,
&&
|A_1\rangle=|C\rangle,
&&
|S_4\rangle=-{\textstyle\frac{1}{2}}|\alpha\rangle,
&&
|\lambda_1\rangle=|\epsilon\rangle.
\end{align}
As to the Lagrange multipliers present in (\ref{action}), they show up as the $b^{+}$--independent parts
of the states
$|A_2\rangle$,  $|A_3\rangle$
\begin{align}
&
|A_2\rangle=|\lambda_2\rangle+|A_2'\rangle,
&&
b|\lambda_2\rangle=0,
&&
b|A_2'\rangle\neq0,
\label{A2}
\\[2pt]
&
|A_3\rangle=|\lambda\rangle+|A_3'\rangle,
&&
b|\lambda\rangle=0,
&&
b|A_3'\rangle\neq0.
\label{A3}
\end{align}
The states $|\lambda\rangle$, $|\lambda_2\rangle$ correspond to the fields
$\lambda^{(s-2)}$, $\lambda^{(s-4)}$ in (\ref{action}),
respectively.

After imposing the gauge fixing conditions and making proper use of the
decompositions (\ref{Psi-}), (\ref{A2}), (\ref{A3}) one can bring
the Lagrangian (\ref{Qaction}) to the form
\begin{eqnarray}
-{\cal{}L}&=&
\langle\varphi|
\bigl\{
l_0|\varphi\rangle-l_1^+|C\rangle-l_2^+|\lambda_2\rangle
\bigr\}
-
\langle{}D|
\bigl\{
l_0|D\rangle-l_1|C\rangle+|\lambda_2\rangle-l_2^+|\lambda\rangle
\bigr\}
\nonumber
\\
&&
{}+
{\textstyle\frac{1}{2}}\langle\alpha|
\bigl\{
l_0|\lambda_3\rangle-l_1|\lambda_2\rangle+l_1^+|\lambda\rangle
\bigr\}
-
\langle{}C|
\bigl\{
l_1|\varphi\rangle-l_1^+|D\rangle-|C\rangle-l_2^+|\lambda_3\rangle
\bigr\}
\nonumber
\\
&&
{}-
\langle\lambda_2|
\bigl\{
l_2|\varphi\rangle+|D\rangle
+{\textstyle\frac{1}{2}}l_1^+|\alpha\rangle-l_2^+|s_5\rangle
\bigr\}
+
\langle\lambda|
\bigl\{
l_2|D\rangle+{\textstyle\frac{1}{2}}l_1|\alpha\rangle+|s_5\rangle+l_2^+|s_6\rangle
\bigr\}
\nonumber
\\
&&
{}+
\langle{}A_2'|K
\bigl\{b^+|s_5\rangle+L_2^+|S_5'\rangle
\bigr\}
+
\langle{}A_3'|K
\bigl\{
|S_5'\rangle+b^+|s_6\rangle+L_2^+|S_6'\rangle
\bigr\}
\label{Qact1}
\\
&&
{}-
\langle{}S_3|K
\bigl\{
-{\textstyle\frac{1}{2}}l_0|\alpha\rangle-l_2|C\rangle-L_2^+|A_4\rangle
\bigr\}
-
\langle{}S_5|K
\bigl\{
l_0|S_5\rangle-L_2|A_2\rangle-|A_3\rangle+l_1^+|A_4\rangle
\bigr\}
\nonumber
\\
&&{}
+
\langle{}S_6|K
\bigl\{
l_0|S_6\rangle+L_2|A_3\rangle-l_1|A_4\rangle
\bigr\}
-
\langle{}A_4|K
\bigl\{
l_1|S_5\rangle-L_2|S_3\rangle
+l_1^+|S_6\rangle+|A_4\rangle
\bigr\}
\nonumber
,
\end{eqnarray}
where the states $|S_5\rangle$, $|S_6\rangle$ were decomposed
into the $b^+$--independent and the $b^+$--dependent parts
\begin{align}
&
|S_5\rangle=|s_5\rangle+|S_5'\rangle,
&&
b_2|s_5\rangle=0,
&&
b_2|S_5'\rangle\neq0,
\nonumber\\[2pt]
&
|S_6\rangle=|s_6\rangle+|S_6'\rangle,
&&
b_2|s_6\rangle=0,
&&
b_2|S_6'\rangle\neq0
.
\end{align}

Variation of the action functional corresponding to the Lagrangian density
(\ref{Qact1}) with respect to $\langle{}A_2'|$ yields the equation of motion
\be
|S_5\rangle=0.
\ee
Substituting this back into the action and varying it with respect to
$\langle{}A_3'|$ one finds
\be
|S_6\rangle=0.
\ee
Being substituted into the action, the latter
removes the states $\langle{}A_2'|$, $\langle{}A_3'|$ from the consideration.
Finally, the equation of motion for $\langle{}A_4|$ allows one to express
$|A_4\rangle$ in terms of $L_2|S_3\rangle$ and to get the reduced action functional
\begin{eqnarray}
-{\cal{}L}&=&
\langle\varphi|
\bigl\{
l_0|\varphi\rangle-l_1^+|C\rangle-l_2^+|\lambda_2\rangle
\bigr\}
-
\langle{}D|
\bigl\{
l_0|D\rangle-l_1|C\rangle+|\lambda_2\rangle-l_2^+|\lambda\rangle
\bigr\}
\nonumber
\\
&&
{}+
{\textstyle\frac{1}{2}}\langle\alpha|
\bigl\{
l_0|\lambda_3\rangle-l_1|\lambda_2\rangle+l_1^+|\lambda\rangle
\bigr\}
-
\langle{}C|
\bigl\{
l_1|\varphi\rangle-l_1^+|D\rangle-|C\rangle-l_2^+|\lambda_3\rangle
\bigr\}
\nonumber
\\
&&
{}-
\langle\lambda_2|
\bigl\{
l_2|\varphi\rangle+|D\rangle
+{\textstyle\frac{1}{2}}l_1^+|\alpha\rangle
\bigr\}
+
\langle\lambda|
\bigl\{
l_2|D\rangle+{\textstyle\frac{1}{2}}l_1|\alpha\rangle
\bigr\}
\nonumber
\\
&&{}
-
\langle{}\lambda_3|K
\bigl\{
-{\textstyle\frac{1}{2}}l_0|\alpha\rangle
-l_2|C\rangle-L_2^+L_2|S_3\rangle
\bigr\}
+
\langle{}S_3'|K
\bigl\{
L_2^+L_2|S_3\rangle
\bigr\}
\label{Qact2}
,
\end{eqnarray}
were we implemented the decomposition
\begin{align}
&
|S_3\rangle=|\lambda_3\rangle+|S_3'\rangle,
&&
b|\lambda_3\rangle=0,
&&
b|S_3'\rangle\neq0.
\label{S3}
\end{align}

At this point we take into account the properties of the operator
$K$ (see e.g. \cite{bpt}) and rewrite
the last term in the Lagrangian (\ref{Qact2}) in the form
$\langle{}S_3'|L_2^+ K\Bigr\{L_2|S_3'\rangle+l_2|\lambda_3\rangle\Bigl\}$.
Thus, on-shell the state $L_2|S_3'\rangle$ is the same as $-l_2|\lambda_3\rangle$.
After the field redefinition
\begin{eqnarray}
|\lambda\rangle
\to
|\lambda\rangle+l_1|\lambda_3\rangle
\qquad
|\lambda_2\rangle
\to
|\lambda_2\rangle+l_1^+|\lambda_3\rangle
\qquad
|C\rangle
\to
|C\rangle-l_2|\lambda_3\rangle
\end{eqnarray}
one gets
\be
|\lambda_3\rangle=0
\ee as the equation of motion, while the Lagrangian (\ref{Qact2}) simplifies to
\begin{eqnarray}\label{bact}
-{\cal{}L}&=&
\langle\phi|
\bigl\{
l_0|\phi\rangle-l_1^+|C\rangle-l_2^+|\lambda_2\rangle
\bigr\}
-
\langle{}D|
\bigl\{
l_0|D\rangle-l_1|C\rangle+|\lambda_2\rangle-l_2^+|\lambda\rangle
\bigr\}
\nonumber
\\
&&{}
-
\langle{}C|
\bigl\{
l_1|\phi\rangle-l_1^+|D\rangle-|C\rangle
\bigr\}
-
{\textstyle\frac{1}{2}}\langle\alpha|
\bigl\{
l_1|\lambda_2\rangle-l_1^+|\lambda\rangle
\bigr\}
\nonumber
\\
&&
{}-
\langle\lambda_2|
\bigl\{
l_2|\phi\rangle+|D\rangle+{\textstyle\frac{1}{2}}l_1^+|\alpha\rangle
\bigr\}
+
\langle\lambda|
\bigl\{
l_2|D\rangle+{\textstyle\frac{1}{2}}l_1|\alpha\rangle
\bigr\}.
\end{eqnarray}
This is precisely the unconstrained Lagrangian (\ref{action}) written in the notation adopted to the
BRST method. The corresponding gauge transformation reads
\bea
\label{GTrans}
&&
\delta|\phi\rangle=l_1^+|\varepsilon\rangle,
\qquad
\delta|D\rangle=l_1|\varepsilon\rangle,
\qquad
\delta|\alpha\rangle=-2l_2|\varepsilon\rangle,
\qquad
\delta|C\rangle=l_0|\varepsilon\rangle.
\eea

Thus, a link between the unconstrained Lagrangian formulation (\ref{action})
and the BRST formulation (\ref{Qaction}) is established via the
partial gauge fixing followed by on--shell elimination of
some of the auxiliary fields.
Notice that
making further steps in this direction, i.e. removing the states
$|C\rangle$ and $|D\rangle$,
one would generate the higher derivative terms in the action
and obtain the Francia--Sagnotti formulation\footnote{As the Lagrange multiplier
$\l^{(s-4)}$ is invariant under the gauge transformation, a field redefinition is needed
in order to relate it with the Lagrange multiplier entering the Francia-Sagnotti formulation.} \cite{fs2}.
The subsequent gauging away of $|\alpha\rangle$ would result in the Fronsdal
model \cite{fr}.

Thus, (\ref{action}) can be viewed
as the {\it minimal} unconstrained Lagrangian formulation for a massless spin--s boson compatible with the
standard requirements imposed on a classical field theory like locality, the absence of higher
derivatives terms etc.

Concluding this section let us discuss a simple truncation of the BRST method, which results
in the same unconstrained equations (\ref{unc}). Consider the operator
\be
g_0=-\frac{1}{2}(a_\m a^{\m + }+a^{+}_\mu a^\mu),
\ee
which forms a closed algebra together with those in
(\ref{oper}). Treating the operators as first class constraints one can associate with
$l_0$, $l_1$, $l_1^{+}$, $l_2$ ,  $l_2^{+}$, $g_0$ the ghost pairs
$(\eta_0,{\cal{}P}_0)$, $(\eta_1^+,{\cal{}P}_1)$, $(\eta_1,{\cal{}P}_1^+)$, $(\eta_2^+,{\cal{}P}_2)$,
$(\eta_2,{\cal{}P}_2^+)$, $(\eta_G,{\cal{}P}_G)$, respectively, and construct the Hermitian and
nilpotent BRST charge
\begin{eqnarray}\label{charge}
Q
&=&
\eta_0l_0+\eta_1^+l_1+\eta_1l_1^+
+\eta_2^+l_2
+\eta_2l_2^+
+\eta_{G}g_0
-\eta_1^+\eta_1{\cal{}P}_0
-\eta_2^+\eta_2{\cal{}P}_G
\nonumber
\\&&
{}
+(\eta_G\eta_1^++\eta_2^+\eta_1){\cal{}P}_1
+(\eta_1\eta_G+\eta_1^+\eta_2){\cal{}P}_1^+
+2\eta_G\eta_2^+{\cal{}P}_2
+2\eta_2\eta_G{\cal{}P}_2^+.
\label{BRST'}
\end{eqnarray}
Here we assume that the ghost operators satisfy the conditions
(\ref{comgh}), (\ref{ghvac}), while
$\eta_G$, $\mathcal{P}_G$ act on the ghost vacuum according to the rule
\begin{eqnarray}
\{\eta_G,\mathcal{P}_G\}=1,
\qquad
\eta_G|0\rangle=0,
\qquad
\mathcal{P}_G|0\rangle\neq0.
\end{eqnarray}

An attempt to relate cohomologies of the BRST charge (\ref{charge}) with irreducible representations of
the Poincar\'e group faces the problem, as the operator $g_0$ is strictly positive
and can not be regarded as providing a physical state condition \cite{pt}. A possible way out was proposed in
\cite{pt} (see also \cite{bpt}). It consists in extending the original Fock space by an extra oscillator $b$ and modifying $g_0$
and other constraints in a proper way. Being perfectly consistent in itself, this recipe, however,
leads to complications. In particular, on anti de Sitter background one has to deal with a nonlinear algebra of
constraints \cite{bpt}, the conventional scalar product is to be modified in order to make the corresponding
BRST charge hermitian etc. The growth of auxiliary fields with the value of spin intrinsic to the BRST description of
higher spin fields is also a consequence of introducing $b$ and $b^{+}$.

Notice that another possibility to cure the above mentioned problem is to impose
additional restrictions that will implement
a consistent truncation of the BRST charge (\ref{charge}) and its cohomologies. Indeed, consider
the following {\it constraints} on the physical states $|\Psi\rangle$ and gauge parameters $|\Lambda\rangle$
\begin{eqnarray}
\eta_2|\Psi\rangle=\eta_G|\Psi\rangle=0, \quad
\eta_2|\Lambda\rangle=\eta_G|\Lambda\rangle=0.
\end{eqnarray}
Out of twenty states originally present in the cohomology the
BRST operator (\ref{charge}) this truncation selects four
\be
|\Psi\rangle=
|\phi\rangle
+\eta_1^+{\cal{}P}_1^+|D\rangle
-\eta_2^+{\cal{}P}_1^+{\textstyle\frac{1}{2}}|\alpha\rangle
+\eta_0{\cal{}P}_1^+|C\rangle.
\ee
The gauge parameter now reads
\be
|\Lambda\rangle=
{\cal{}P}_1^+|\epsilon\rangle.
\ee
Although the truncated BRST charge
\begin{eqnarray}\label{trunc}
Q
&=&
\eta_0l_0+\eta_1^+l_1+\eta_1l_1^+
+\eta_2^+l_2
-\eta_1^+\eta_1{\cal{}P}_0
+\eta_2^+\eta_1{\cal{}P}_1
\end{eqnarray}
is no longer Hermitian with respect to a conventional scalar product,
its continues to be nilpotent and can be used to determine the
dynamics of the reduced physical states. In particular, the condition
\be
Q|\Psi\rangle=0
\ee
gives the set of equations
\bea
&&
l_0|\phi\rangle-l_1^+|C\rangle
=0, \qquad \qquad
l_1|\phi\rangle-l_1^+|D\rangle
-|C\rangle
=0,
\nonumber\\[2pt]
&&
l_2|\phi\rangle+|D\rangle
 +{\textstyle\frac{1}{2}}l_1^+|\alpha\rangle
=0, \quad ~
l_2|D\rangle+{\textstyle\frac{1}{2}}l_1|\alpha\rangle
=0,
\nonumber\\[2pt]
&&
l_0|D\rangle-l_1|C\rangle
 =0, \qquad \qquad
{\textstyle\frac{1}{2}}l_0|\alpha\rangle+l_2|C\rangle
=0,
\eea
which are those in (\ref{unc}) and differential consequences thereof.
The truncated gauge transformation $\delta|\Psi\rangle=Q|\Lambda\rangle$
correctly reproduces (\ref{alp}) and (\ref{other}).

Notice that imposing one more constraint
${\cal{}P}_2 |\Psi\rangle=0$ on the physical states $|\Psi\rangle$, one does not alter the BRST charge (\ref{trunc}) but
removes the compensator $\a^{(s-3)}$ from the consideration. The resulting constrained formulation,
which involves a traceless gauge parameter, was previously discussed in \cite{g}, \cite{fpt}.

\vspace{0.5cm}
\noindent

{\bf 6. Coupling unconstrained higher spin fields to (anti) de Sitter background}\\

As is well known, consistent gravitational interactions of massless higher spin fields can
be formulated only in anti de Sitter space \cite{vas},\cite{vas1},\cite{vas2},\cite{fv},\cite{fv1}.
By this reason it is important to analyse
whether the unconstrained Lagrangian formulations for free massless higher spin
fields constructed in this work can be consistently coupled to anti de Sitter background.

\vspace{0.5cm}
\noindent

{\it 6.1. Bosonic higher spin fields on (anti) de Sitter background}\\

When coupling unconstrained bosonic higher spin fields to (anti) de Sitter background, it is natural
to start with the triplet equations. The corresponding results are available due to the work of
\cite{st}.
Following ref. \cite{st}, we minimally extend
the gauge transformation laws of the
fields $\phi^{(s)}$ and $D^{(s-2)}$
\be\label{ctr}
\d \phi^{(s)}=(y\nabla)\e^{(s-1)}, \quad
\quad \d D^{(s-2)}=
(\pi\nabla)\e^{(s-1)},
\ee
where $\nabla_\m$ is the covariant derivative operator
\footnote{In order to put bulky expressions involving symmetrized vector indices in a compact form,
we introduce an auxiliary variable $y^\m$, the corresponding derivative
$\pi_\m=\frac{\partial}{\partial y^\m}$, and use the notation $y^2=g_{\m\n}y^\m y^\n$, $\pi^2=g^{\m\n}\pi_\m \pi_\n$,
where $g_{\m\n}$ is a background metric.
The operators $y^\m$ and $\pi_\m$ are taken to be inert under
the action of the general coordinate group and commuting with the covariant derivative operator $\nabla_\m$.
Although the object $\phi^{(s)}(x,y)=\phi_{\m_1 \dots \m_s} (x)~ y^{\m_1} \dots y^{\m_s}$ appearing at an
intermediate stage of calculation is not invariant under the general coordinate group, the final action functional does not
involve $y^\m$ and, as thus, respects the general coordinate invariance. When using this formalism on (anti)
de Sitter background, where $R_{\a\b\m\n}= r (g_{\a\m} g_{\b\n}-g_{\a\n} g_{\b \m} )$,
particularly useful relations are $[\nabla_\m,\nabla_\n] \phi^{(s)}= -y^\a \pi_\b {R^\b}_{\a\m\n} \phi^{(s)}$,
$[(\pi\nabla),(y \nabla)]\phi^{(s)}=\nabla^2 \phi^{(s)}-ry^2\pi^2 \phi^{(s)}+rs[s+N-2]\phi^{(s)}$.}
, as well as the equation of motion, which determines $C^{(s-1)}$
\be\label{backc}
C^{(s-1)}-(\pi\nabla)\phi^{(s)}
+(y\nabla) D^{(s-2)}=0.
\ee
The transformation law for the latter then follows from (\ref{ctr})
\be\label{ctr1}
\d C^{(s-1)}=[(\pi\nabla),(y\nabla)] \e^{(s-1)}=\nabla^2 \e^{(s-1)}+r(s-1)(s+N-3)\e^{(s-1)}-ry^2\pi^2 \e^{(s-1)},
\ee
where $r$ is the constant entering the curvature tensor of
(anti) de Sitter space--time
\be\label{curvat}
R_{\a\b\m\n}= r (g_{\a\m} g_{\b\n}-g_{\a\n} g_{\b \m} ).
\ee

Variation of the remaining triplet equations minimally coupled to background metric
under the gauge transformations (\ref{ctr}), (\ref{ctr1}) then yields the terms, which should be
taken into account in order to get the fully gauge invariant description
\bea\label{backp}
&&
\left( \nabla^2 -r y^2 \pi^2 +r\left[(s-2)(s+N-3)-s \right]\right)\phi^{(s)} -(y\nabla) C^{(s-1)}
+4ry^2 D^{(s-2)}=0,
\\[2pt]\label{backd}
&&
\left( \nabla^2 -r y^2 \pi^2 +r\left[s(s+N-2)+6 \right] \right) D^{(s-2)} -(\pi\nabla) C^{(s-1)}
-4r\pi^2 \phi^{(s)}=0.
\eea

At this point we introduce the compensator $\a^{(s-3)}$, which
transforms under the gauge transformation according to the rule
\be
\d\a^{(s-3)}=\frac{\pi^2}{2} \e^{(s-1)}
\ee
and impose the gauge invariant constraints
\be\label{backcon}
D^{(s-2)}-\frac{\pi^2}{2} \phi^{(s)} +(y\nabla)\a^{(s-3)}=0, \qquad
\frac{\pi^2}{2} D^{(s-2)} -(\pi\nabla)\a^{(s-3)}=0.
\ee
Notice that, after introducing these constraints, the triplet equation (\ref{backd}) becomes the
differential consequence
of (\ref{backc}), (\ref{backp}) and (\ref{backcon}). Besides, one can readily verify that, after gauging away
the compensator $\a^{(s-3)}$ and eliminating $C^{(s-1)}$, $D^{(s-2)}$ with the use of their
equations of motion, one is left with an $N$--dimensional generalization \cite{bpt} of the Fronsdal equation
\cite{fads}, which describes a massless spin--s boson coupled to (anti) de Sitter background.

An unconstrained Lagrangian formulation reproducing the formalism outlined above is the sum of
the triplet Lagrangian \cite{st} and the constraint terms, which are enforced
with the use of two Lagrange multipliers $\lambda^{(s-2)}$, $\lambda^{(s-4)}$
\bea\label{backact}
&&
S=-{(-1)}^s \int d^N x \Bigl\{ \textstyle{\frac{1}{2}} \phi_{\m_1 \dots \m_s} \nabla^2 {}  \phi^{\m_1 \dots \m_s}-
s\phi_{\m_1 \dots \m_s}  \nabla^{\m_1} C^{\m_2 \dots \m_s} -
\textstyle{\frac{1}{2}} s C_{\m_1 \dots \m_{s-1}} C^{\m_1 \dots \m_{s-1}} -
\nonumber\\[2pt]
&&
\quad \qquad -s(s-1)C_{\m_1 \dots \m_{s-1}}\nabla^{\m_1} D^{\m_2 \dots \m_{s-1}} -
\textstyle{\frac{1}{2}} s(s-1) D_{\m_1 \dots \m_{s-2}} \nabla^2 D^{\m_1 \dots \m_{s-2}}+
\nonumber\\[2pt]
&&
\quad \qquad  +\textstyle{\frac{1}{2}}r[(s-2)(s+N-3)-s]\phi_{\m_1 \dots \m_s}\phi^{\m_1 \dots \m_s}
-\textstyle{\frac{1}{2}}s(s-1)r {\phi^\n}_{\n\m_1 \dots \m_{s-2}}{{\phi^\s}_\s}^{\m_1 \dots \m_{s-2}}+
\nonumber\\[2pt]
&&
\quad \qquad  +4s(s-1)r{\phi^\n}_{\n\m_1 \dots \m_{s-2}} D^{\m_1 \dots \m_{s-2}}+
\textstyle{\frac{1}{2}}s(s-1)(s-2)(s-3)r {D^\n}_{\n\m_1 \dots \m_{s-4}}{{D^\s}_\s}^{\m_1 \dots \m_{s-4}}-
\nonumber\\[2pt]
&&
\quad \qquad  -\textstyle{\frac{1}{2}} s(s-1)r[s(s+N-2)+6]D_{\m_1 \dots \m_{s-2}} D^{\m_1 \dots \m_{s-2}}+
\nonumber\\[2pt]
&&
\quad \qquad  +\lambda^{\mu_1\ldots\mu_{s-2}}\left(\textstyle{\frac{1}{2}} {\phi^\n}_{\n \mu_1 \dots \m_{s-2}}
-D_{\m_1 \dots\m_{s-2}}
-\textstyle{\frac{1}{2}} (s-2)\nabla_{\m_1}\alpha_{\m_2 \dots \m_{s-2}}\right)+
\nonumber\\[2pt]
&&
\quad \qquad  +
\lambda^{\m_1 \dots \m_{s-4}} \left({D^\n}_{\n \m_1 \dots \m_{s-4}}-\nabla^\n \a_{\n \m_1 \dots \m_{s-4}}\right)
\Bigr\}.
\eea
The equations given above follow from the Lagrangian after the field redefinition
\be
s C^{(s-1)} \rightarrow C^{(s-1)}, \quad s(s-1) D^{(s-2)} \rightarrow D^{(s-2)}, \quad
\frac{s(s-1)(s-2)}{2} \a^{(s-3)} \rightarrow \a^{(s-3)}.
\ee
The action (\ref{backact}) is invariant under the gauge transformation
\bea
&&
\delta \phi_{\mu_1 \dots \mu_s}=
\nabla_{\mu_1} \epsilon_{\mu_2 \dots \mu_s}+\dots, \quad
\delta D_{\mu_1 \dots\mu_{s-2}}=
\nabla^\nu \epsilon_{\nu\mu_1\dots\mu_{s-2}}, \quad
\delta \alpha_{\mu_1\dots\mu_{s-3}}=
{\epsilon^\nu}_{\nu\mu_1\dots\mu_{s-3}},
\nonumber\\[2pt]
&&
\delta C_{\mu_1\ldots\mu_{s-1}}=
[\nabla^2+r(s-1)(s+N-3)]\epsilon_{\mu_1\dots\mu_{s-1}}
-2r(g_{\mu_1\mu_2}\epsilon^\nu{}_{\nu\mu_3\ldots\mu_{s-1}}
+\dots).
\eea

Finally, applying the same arguments as in flat space one can demonstrate that the Lagrange multipliers
$\lambda^{(s-2)}$, $\lambda^{(s-4)}$
vanish on-shell and do not alter eqs. (\ref{backc}), (\ref{backp}), (\ref{backcon}). Notice that as in
flat space the resulting formulation is local, free from higher derivative terms and involves the same number
of auxiliary fields for any value of spin. Eliminating
$C^{(s-1)}$, $D^{(s-2)}$ from the action (\ref{backact}) and redefining $\lambda^{(s-4)}$ one obtains
the higher derivative formulation of \cite{fms}.

\vspace{0.5cm}
\noindent

{\it 6.2 Fermionic higher spin fields on (anti) de Sitter background}\\

When coupling the fermionic triplet to (anti) de Sitter background, one encounters
certain difficulties \cite{st} and a corresponding Lagrangian formulation
has not been constructed yet. The point is that, being properly combined, the triplet equations
in (anti) de Sitter space yield one more restriction
\be\label{pr}
C^{(n-1)}-(\g\pi)\Psi^{(n)}=0,
\ee
thus indicating inconsistency of the coupling \cite{st}.
Beautifully enough, comparing (\ref{pr}) with (\ref{uncf}) one observes that this is precisely
one of the constraints used in our unconstrained description of a massless spin-s fermion in flat space
modulo the term including the compensator $\a^{(n-2)}$. Thus, a seemingly problematic
point of the triplet formulation turns into advantage when one uses it for
constructing an unconstrained Lagrangian formulation for a massless spin-s
fermion in (anti) de Sitter space.

Below we construct
a generalization of eqs. (\ref{uncf}) and their symmetries
to the case of (anti) de Sitter
space\footnote{In this section we use the following notation. The conventional Dirac matrices are
contracted with the vielbein such that $\g_\m \g_\n+\g_\n \g_\m=g_{\m\n}$, where $g_{\m\n}$ is
the background metric. As the vielbein is covariantly constant, $\g_\m$ commute with
the covariant derivative $\nabla_\n$. The auxiliary variables $y^\m$ do not transform under
the general coordinate transformations and
commute with the covariant derivative $\nabla_\n$. Contractions of $y^\m$ and
$\pi_\m=\frac{\partial}{\partial y^\m}$ with the $\g$--matrices are implemented
in the following way  $(\g y)=\g_\m y^\m$, $(\g\pi)=\g_\m g^{\m\n} \pi_\n$.
It is assumed that each tensor field appearing in
this section carries the Dirac spinor index $A$, which is suppressed throughout.
Given a spin--tensor $\vf^{(n)}$ in (anti) de Sitter space, the commutator of two covariant derivatives
reads
$[\nabla_\m,\nabla_\n] \vf^{(n)}=\frac{1}{4}r {(\g_\m \g_\n -\g_\n \g_\m)} \vf^{(n)}-y^\a \pi_\b
{R^\b}_{\a\m\n}\vf^{(n)}$, with ${R^\b}_{\a\m\n}$ from (\ref{curvat}). For practical calculations
particularly useful formulae are
$[(\g\nabla), (y\nabla)] \vf^{(n)}=-r y^2 (\g\pi) \vf^{(n)}+\frac{1}{2}r(N+2n-1)(\g y) \vf^{(n)}$,
$[(\pi\nabla), (y\nabla)] \vf^{(n)}=\nabla^2 \vf^{(n)}-\frac{1}{2}r(\g y) (\g\pi)\vf^{(n)}-ry^2 \pi^2 \vf^{(n)}
+rn(N+n-\frac{3}{2})\vf^{(n)} $,
$[(\g\nabla), (\pi\nabla)] \vf^{(n)}=r(\g y) \pi^2 \vf^{(n)}-\frac{1}{2}r(N+2n-3)(\g \pi) \vf^{(n)}$,
$(\g\nabla)(\g\nabla)\vf^{(n)}=\nabla^2 \vf^{(n)}+r(\g y)(\g\pi)\vf^{(n)}-\frac{1}{4}r(N^2-N+4n)\vf^{(n)}$,
$\{ (\g\pi), (\g y)\}\vf^{(n)}=(N+2n)\vf^{(n)}$, where $N$ stands for the dimension of space--time.}.
The knowledge of specific gauge invariant
combinations (constraints) will allow us to properly extend the triplet equations.
In contrast to flat space, the generalizations will explicitly involve
the compensator field $\a^{(n-2)}$ due to the
effects of non minimal interaction. Then we construct a Lagrangian density describing
the generalized triplet and the gauge invariant constraints, which altogether
provide a simple unconstrained formulation for a massless spin-s fermion
in (anti) de Sitter space.

Guided by the analysis of Fang and Fronsdal \cite{fang}, we take the gauge transformation law for
$\Psi^{(n)}$ in the form
\be\label{adstran1}
\d \Psi^{(n)}=(y\nabla) \e^{(n-1)}+\frac{i}{2}\sqrt{r} (y\g) \e^{(n-1)}.
\ee
Transformation laws of other fields
\bea
&&
\d C^{(n-1)}=(\g\nabla) \e^{(n-1)}+\frac{i}{2}\sqrt{r}(N+2n-2))\e^{(n-1)},
\nonumber
\eea
\bea\label{adstran2}
&&
\d D^{(n-2)}=(\pi\nabla) \e^{(n-1)}, \qquad \d \a^{(n-2)}=(\g \pi) \e^{(n-1)},
\eea
and corrections in (\ref{uncf}) caused by the effects of non minimal interaction
\bea\label{fads}
&&
(\g \nabla)\Psi^{(n)} -(y\nabla) C^{(n-1)}+\frac{i}{2}\sqrt{r}(N+2n-4)\Psi^{(n)}+\frac{i}{2}\sqrt{r}(\g y)
(\g\pi)\Psi^{(n)}
+\frac{3}{4}ry^2 \a^{(n-2)}-
\nonumber\\[2pt]
&& - \frac{i}{2}\sqrt{r} (\g y)(y\nabla) \a^{(n-2)} =0,
\\[2pt]\label{fads1}
&&
C^{(n-1)}-(\g \pi) \Psi^{(n)} +(y\nabla)\a^{(n-2)}-\frac{i}{2}\sqrt{r}(\g y)\a^{(n-2)}=0,
\\[2pt]\label{fads2}
&&
D^{(n-2)}-\frac{1}{2} (\g\pi)C^{(n-1)}-\frac{1}{2}(\g \nabla) \a^{(n-2)}+\frac{i}{4}\sqrt{r}(N+2n-2)\a^{(n-2)}=0,
\\[2pt]\label{fads3}
&&
(\g\pi) D^{(n-2)}-(\pi \nabla)\a^{(n-2)}=0
\eea
follow from the requirement that the formalism is to reproduce
the Fang--Fronsdal equations \cite{fang} when $\a^{(n-2)}$ is
gauged away and $C^{(n-1)}$, $D^{(n-2)}$ are eliminated with the use of their equations of motion.

A generalization of the triplet equations (\ref{fsa}) to (anti) de Sitter space is found with the use of
the gauge transformations (\ref{adstran1}), (\ref{adstran2})
\bea\label{fads4}
&&
(\g\nabla) C^{(n-1)}-(\pi\nabla)\Psi^{(n)}+(y\nabla) D^{(n-2)}-\frac{i}{2}\sqrt{r}(N+2n-3)C^{(n-1)}+
\nonumber\\[2pt]
&&
+\frac{i}{2}\sqrt{r}(\g y) D^{(n-2)}-\frac{3}{2}r (\g y) \a^{(n-2)} -r y^2 (\g \pi) \a^{(n-2)}=0,
\\[2pt]\label{fads5}
&&
(\g\nabla)D^{(n-2)}-(\pi\nabla) C^{(n-1)}+\frac{i}{2}\sqrt{r}(N+2n-2)D^{(n-2)}+\frac{1}{2}r(N+2n-5)\a^{(n-2)}-
\nonumber\\[2pt]
&&
-r(\g y) (\g\pi)\a^{(n-2)}=0.
\eea
As in flat space these relations prove to be the differential consequences of (\ref{fads})--(\ref{fads3}).

The construction of an action functional reproducing (\ref{fads})--(\ref{fads5}) proves to be
more complicated than in flat space. Starting from the generalized triplet equations (\ref{fads}),
(\ref{fads4}), {}(\ref{fads5}), one finds that it is problematic to construct a {\it real} action functional
unless one uses the constraint (\ref{fads2}). Indeed, the term ${\bar D}_{\m_1 \dots \m_{n-2}}
\g^\n {C_\n}^{\m_1 \dots \m_{n-2}}$, which is to be included into the action so as to make it real, spoils the
triplet equations. A natural way out is to extend $\g^\n {C_\n}^{\m_1 \dots \m_{n-2}}$
to the full expression, which stands on the l.h.s. of (\ref{fads2}). Where appropriate, complex conjugates
should be added to the action. Then the undesirable
contribution into the triplet equations turns into the constraint and, as thus, becomes harmless. Finally, one demands
the action to be gauge invariant and finds further contributions, which prove to be
quadratic in the compensator $\a^{(n-2)}$
\bea\label{fermact}
&&
S=\int d^N x \sqrt{-g} \Bigl\{ \bar\Psi_{\m_1 \dots\m_n}\left[ i~ \slash\!\!\!\!\nabla~ \Psi^{\m_1 \dots\m_n}
-in\nabla^{\m_1} C^{\m_2 \dots \m_n}
-\textstyle{\frac{1}{2}}(N+2(n-2))\sqrt{r}\Psi^{\m_1 \dots\m_n}-
\right.
\nonumber\\[2pt]
&&
-\textstyle{\frac{1}{2}}n\sqrt{r}\g^{\m_1} \g^\n {\Psi_\n}^{\m_2 \dots\m_n}+
\textstyle{\frac{3}{4}}in(n-1)r g^{\m_1 \m_2} \a^{\m_3 \dots\m_n}+
\textstyle{\frac{1}{2}}n(n-1)\sqrt{r}\g^{\m_1} \nabla^{\m_2} \a^{\m_3 \dots\m_n}+
\nonumber
\eea
\bea
&&
\left.
+in\g^{\m_1}\lambda^{\m_2 \dots\m_n}\right]
+n{\bar C}_{\m_1 \dots\m_{n-1}}\left[
i~ \slash\!\!\!\!\nabla~ C^{\m_1 \dots\m_{n-1}}-i \nabla^\n {\Psi_\n}^{\m_1 \dots\m_{n-1}}
+i(n-1)\nabla^{\m_1} D^{\m_2 \dots \m_{n-1}}+
\right.
\nonumber\\[2pt]
&&
+\textstyle{\frac{1}{2}}(N+2(n-1)-1)\sqrt{r} C^{\m_1 \dots\m_{n-1}}
-\textstyle{\frac{1}{2}}(n-1)\sqrt{r}\g^{\m_1}D^{\m_2 \dots\m_{n-1}}
-\textstyle{\frac{3}{2}}i(n-1)r \g^{\m_1} \a^{\m_2 \dots \m_{n-1}}-
\nonumber\\[2pt]
&&
\left.
-i(n-1)(n-2)r g^{\m_1\m_2} \g^\n {\a_\n}^{\m_3 \dots \m_{n-1}}
+\textstyle{\frac{1}{2}}i(n-1)\g^{\m_1}\lambda^{\m_2 \dots\m_{n-1}}
-i\lambda^{\m_1 \dots\m_{n-1}}
\right]+
\nonumber\\[2pt]
&&
+n(n-1){\bar D}_{\m_1 \dots\m_{n-2}}\left[-i ~\slash\!\!\!\!\nabla~ D^{\m_1 \dots\m_{n-2}}
+i\nabla^\n {C_\n}^{\m_1 \dots\m_{n-2}}-\textstyle{\frac{1}{2}}\sqrt{r}\g^\n {C_\n}^{\m_1 \dots\m_{n-2}}
-\textstyle{\frac{1}{2}}\sqrt{r}~\slash\!\!\!\!\nabla \a^{\m_1 \dots\m_{n-2}}
\right.
\nonumber\\[2pt]
&&
+\textstyle{\frac{1}{2}}(N+2n)\sqrt{r} D^{\m_1 \dots\m_{n-2}}
-\textstyle{\frac{1}{4}}i(N+2(n-2)-4)r \a^{\m_1 \dots\m_{n-2}}
+i(n-2)r \g^{\m_1} \g^\n {\a_\n}^{\m_2 \dots\m_{n-2}}-
\nonumber\\[2pt]
&&
\left.
-i\lambda^{\m_1 \dots\m_{n-2}}-i(n-2)\g^{\m_1}
\lambda^{\m_2 \dots \m_{n-2}}\right]
+n(n-1){\bar\a}_{\m_1 \dots\m_{n-2}}\left[\textstyle{\frac{1}{2}}\sqrt{r}~\nabla^2 \a^{\m_1 \dots\m_{n-2}}+
\right.
\nonumber\\[2pt]
&&
+
\textstyle{\frac{1}{2}}(n-2)\sqrt{r}~\nabla^{\m_1}\nabla^\n {\a_\n}^{\m_2 \dots\m_{n-2}}
-i(n-2)r \g^{\m_1} \slash\!\!\!\!\nabla \g^\n {\a_\n}^{\m_2 \dots\m_{n-2}}
+\textstyle{\frac{5}{4}}ir ~\slash\!\!\!\!\nabla \a^{\m_1 \dots\m_{n-2}}+
\nonumber\\[2pt]
&&
+\textstyle{\frac{3}{4}}i(n-2)r~ \nabla^{\m_1} \g^\n {\a_\n}^{\m_2 \dots\m_{n-2}}
+\textstyle{\frac{3}{4}}i(n-2)r~ \g^{\m_1} \nabla^\n  {\a_\n}^{\m_2 \dots\m_{n-2}}
-\textstyle{\frac{3}{4}}ir {\Psi_\n}^{\n \m_1 \dots\m_{n-2}}+
\nonumber\\[2pt]
&&
+\textstyle{\frac{1}{2}}n(N+n-\frac{3}{2})r^{3/2}\a^{\m_1 \dots\m_{n-2}}-
\textstyle{\frac{1}{2}}(n-2)(n-3)r^{3/2} g^{\m_1\m_2} {\a_\n}^{\n\m_3 \dots\m_{n-2}}+
\nonumber\\[2pt]
&&
+\textstyle{\frac{1}{2}}(n-2)(N+2(n-2)-\textstyle{\frac{1}{4}}) r^{3/2} \g^{\m_1} \g^\n
{\a_\n}^{\m_2 \dots\m_{n-2}}
+i(n-2)r \g^{\m_1} {C_\n}^{\n \m_2 \dots\m_{n-2}}-
\nonumber\\[2pt]
&&
-\textstyle{\frac{1}{2}}\sqrt{r}~ \nabla^\n \g^\s {\Psi_{\n\s}}^{\m_1 \dots\m_{n-2}}
+\textstyle{\frac{3}{2}}ir~\g^\n
{C_\n}^{\m_1 \dots\m_{n-2}}+\textstyle{\frac{1}{2}}i(N+2(n-2)-1)rD^{\m_1 \dots \m_{n-2}}-
\nonumber\\[2pt]
&&
-i(n-2)r \g^{\m_1}\g^\n {D_\n}^{\m_2 \dots \m_{n-2}}-\textstyle{\frac{1}{4}}i(N+2(n-1))rD^{\m_1 \dots \m_{n-2}}
+\textstyle{\frac{1}{2}}\sqrt{r}~\slash\!\!\!\!\nabla D^{\m_1 \dots \m_{n-2}}+
\nonumber\\[2pt]
&&
+i\nabla^\n
{\lambda_\n}^{\m_1 \dots \m_{n-2}}+\textstyle{\frac{1}{2}}\sqrt{r}\g^{\n}{\lambda_\n}^{\m_1 \dots \m_{n-2}}
-\textstyle{\frac{1}{2}} i~\slash\!\!\!\!\nabla \lambda^{\m_1 \dots \m_{n-2}}
-i(n-2) \nabla^{\m_1} \lambda^{\m_2 \dots \m_{n-2}}-
\nonumber\\[2pt]
&&
\left.
-\textstyle{\frac{1}{4}}(N+2(n-1))\sqrt{r} ~\lambda^{\m_1 \dots \m_{n-2}}\right]
+in {\bar \lambda}_{\m_1 \dots\m_{n-1}} \left[C^{\m_1 \dots\m_{n-1}}-\gamma^\n {\Psi_\n}^{\m_1 \dots \m_{n-1}}+
\right.
\nonumber\\[2pt]
&&
\left.
+(n-1) \nabla^{\m_1} \a^{\m_2 \dots \m_{n-1}}-\textstyle{\frac{1}{2}}i(n-1)\sqrt{r}\g^{\m_1}\a^{\m_2 \dots \m_{n-1}}
\right]+i n(n-1){\bar \lambda}_{\m_1 \dots\m_{n-2}} \left[D^{\m_1 \dots\m_{n-2}}-
\right.
\nonumber\\[2pt]
&&
\left.
-\textstyle{\frac{1}{2}}\gamma^\n {C_\n}^{\m_1 \dots \m_{n-2}}
-\textstyle{\frac{1}{2}} \slash\!\!\!\!\nabla \a^{\m_1 \dots \m_{n-2}}+
\textstyle{\frac{1}{4}}i(N+2(n-1))\sqrt{r} \a^{\m_1 \dots \m_{n-2}}\right]+
\nonumber\\[2pt]
&&
+i n(n-1)(n-2){\bar \lambda}_{\m_1 \dots\m_{n-3}} \left[ \gamma^\n {D_\n}^{\m_1 \dots \m_{n-3}}-
\nabla^\n {\a_\n}^{\m_1 \dots \m_{n-3}}\right]\Bigr\}.
\eea
This action holds invariant under the gauge transformation
\bea
&&
\d \Psi_{\m_1 \dots \m_n}=(\nabla_{\m_1} \e_{\m_2 \dots \m_n}+\dots)+
\frac{1}{2}i\sqrt{r}(\g_{\m_1}\e_{\m_2 \dots \m_n}+\dots),
\nonumber
\eea
\bea
&&
\d C_{\m_1 \dots \m_{n-1}}=\slash\!\!\!\!\nabla~\e_{\m_1 \dots \m_{n-1}} +\frac{1}{2}i\sqrt{r}(N+2(n-1))
\e_{\m_1 \dots \m_{n-1}},
\nonumber\\[2pt]
&&
\d D_{\m_1 \dots \m_{n-2}}=\nabla^\n \e_{\n\m_1 \dots \m_{n-2}}, \qquad
\d \a_{\m_1 \dots \m_{n-2}}=\g^\n
\e_{\n \m_1 \dots \m_{n-2}}.
\eea

Variation of the action with respect to all the fields and Lagrange multipliers gives seven complex equations.
Proceeding in exactly the same way as in sect. 4, one can demonstrate that the Lagrange multipliers vanish
on--shell. Notice that the equation of motion for $\a^{(n-2)}$ involves higher derivative terms. However,
this proves to be the differential consequence of other equations of motion and can be discarded.

Thus, we have demonstrated that the unconstrained description (\ref{fads})--(\ref{fads3})
of a massless spin-s fermion in (anti) de Sitter space admits a consistent Lagrangian formulation.

\vspace{0.5cm}
\noindent

{\bf 7. Conclusion}\\

To summarize, in this paper we have constructed
simple unconstrained Lagrangian formulations for free massless higher spin fields both in
flat space of arbitrary dimension and on (anti) de Sitter background.
The formulations are local, free from higher derivative
terms and use equal number of auxiliary fields for an unconstrained description of any value of spin.
In this setting an irreducible representation of the Poincar\'e group is described
in terms of a quartet of fields.
The models occupy an intermediate
position between the geometric formulations of \cite{fs1}--\cite{fms}
and the BRST models of
\cite{pt}--\cite{bkp} and enjoy all the standard features of a conventional
classical field theory. Our considerations also highlight the important role of the
bosonic and fermionic triplets \cite{fs,fs2} in higher spin gauge theory.

Let us mention a few possible developments of the present work.
First of all, it would be interesting to generalize the present
analysis to the case of mixed--symmetry tensor fields (see e.g.
\cite{bek},
\cite{brink}--\cite{asv}). Then it is tempting to realize similar
mechanism for massive higher spin fields, where the issue of
auxiliary fields becomes much more complicated (see e.g. \cite{mass}
and references therein). Finally, it is interesting to construct
supersymmetric generalizations of the models considered in this
work.

\vspace{0.5cm}

{\bf Acknowledgements}\\

\noindent
We thank A.K.H. Bengtsson, G. Bonelli, N. Boulanger, D. Francia, M. Grigoriev and A. Sagnotti for useful comments.
This research was supported in part by RF Presidential grants
NS-4489.2006.2, MD-8970.2006.2, INTAS grant 03-51-6346, DFG grant 436 RUS 113/669/0-3,
RFBR-DFG grant 06-02-04012
and RFBR grant 06-02-16346.


\begin{thebibliography}{nn}
\bibitem{pt} A. Pashnev, M. Tsulaia, Mod. Phys. Lett. A {\bf 13} (1998) 1853.
\bibitem{bupt} C. Burdik, A. Pashnev, M. Tsulaia, Mod. Phys. Lett. A {\bf 16} (2001) 731.
\bibitem{bpt} I.L. Buchbinder, A. Pashnev, M. Tsulaia, Phys. Lett. B {\bf 523} (2001) 338.
\bibitem{pash} I.L. Buchbinder, A. Pashnev, M. Tsulaia, {\it
Massless higher spin fields in the AdS background and BRST constructions for nonlinear algebras},
arXiv: hep-th/0206026.
\bibitem{bekaert} X. Bekaert, I.L. Buchbinder, A. Pashnev, M. Tsulaia,
Class. Quant. Grav. {\bf 21} (2004) S1457.
\bibitem{bkp} I.L. Buchbinder, V.A. Krykhtin, A.I. Pashnev, Nucl. Phys. B {\bf 711} (2005) 367.
\bibitem{fs1} D. Francia, A. Sagnotti, Phys. Lett. B {\bf 543} (2002) 303.
\bibitem{fs} D. Francia, A. Sagnotti, Class. Quant. Grav. {\bf 20} (2003) S473.
\bibitem{fs2} D. Francia, A. Sagnotti, Phys. Lett. B {\bf 624} (2005) 93.
\bibitem{fs3} D. Francia, A. Sagnotti, J. Phys. Conf. Ser. {\bf 33} (2006) 57.
\bibitem{fms} D. Francia, J. Mourad, A. Sagnotti, {\it Current exchanges and unconstrained higher spins},
arXiv:hep-th/0701163.
\bibitem{boul1} X. Bekaert, N. Boulanger, Phys. Lett. B {\bf 561} (2003) 183.
\bibitem{boul} X. Bekaert, N. Boulanger, Commun. Math. Phys. {\bf 245} (2004) 27.
\bibitem{bek} X. Bekaert, N. Boulanger, {\it Tensor gauge fields in arbitrary representations of GL(D,R):
II. Quadratic actions}, arXiv: hep-th/0606198.
\bibitem{g} G. Barnich, M. Grigoriev, A. Semikhatov, I. Tipunin, Commun. Math. Phys. {\bf 260} (2005)
147.
\bibitem{b2} A.K.H. Bengtsson, J. Math. Phys. {\bf 46} (2005) 042312.
\bibitem{b3} A.K.H. Bengtsson, {\it Structure of higher spin gauge interactions},
arXive: hep-th/0611067.
\bibitem{fr} C. Fronsdal, Phys. Rev. D {\bf 18} (1978) 3624.
\bibitem{ff} J. Fang, C. Fronsdal, Phys. Rev. D {\bf 18} (1978) 3630.
\bibitem{st} A. Sagnotti, M. Tsulaia, Nucl. Phys. B {\bf 682} (2004) 83.
\bibitem{bonelli} G. Bonelli, Nucl. Phys. B {\bf 669} (2003) 159.
\bibitem{lindst} U. Lindstrom, M. Zabzine, Phys. Lett. B {\bf 584} (2004) 178.
\bibitem{v} M.A. Vasiliev, {\it Higher spin gauge theories: star-product and AdS space}, arXiv:hep-th/9910096.
\bibitem{vas} M.A. Vasiliev, Phys. Lett. B {\bf 243} (1990) 378.
\bibitem{vas1} M.A. Vasiliev, Phys. Lett. B {\bf 285} (1992) 225.
\bibitem{vas2} M.A. Vasiliev, Phys. Lett. B {\bf 567} (2003) 139.
\bibitem{r} M.A. Vasiliev, Fortsch. Phys. {\bf 52} (2004) 702.
\bibitem{r1} D. Sorokin, {\it Introduction to the classical theory of higher spins},
arXiv: hep-th/0405069.
\bibitem{r2} X. Bekaert, S. Cnockaert, C. Iazeolla, M.A. Vasiliev, {\it Nonlinear higher spin theories in
varions dimensions}, arXiv: hep-th/0503128.
\bibitem{ost} S. Ouvry, J. Stern, Phys. Lett. B {\bf 177} (1986) 335.
\bibitem{b} A.K.H. Bengtsson, Phys. Lett. B {\bf 182} (1986) 321.
\bibitem{b1} A.K.H. Bengtsson, Class. Quantum Grav. {\bf 5} (1988) 437.
\bibitem{bfpt} I.L. Buchbinder, A. Fotopoulos, A.C. Petkou, M. Tsulaia, Phys. Rev. D {\bf 74} (2006) 105018.
\bibitem{mass} I.L. Buchbinder, V.A. Krykhtin, Nucl. Phys. B {\bf 727} (2005)
537.
\bibitem{mass1}
I.L. Buchbinder, V.A. Krykhtin, L.L. Ryskina, H. Takata, Phys.
Lett. B {\bf 641} (2006) 386.
\bibitem{mass2}
I.L. Buchbinder, P.M. Lavrov, V.A.
Krykhtin, Nucl. Phys. B {\bf 762} (2007) 344.
\bibitem{nonlin} K. Schoutens, A. Servin, P. van Nieuwenhuizen, Commun. Math. Phys. {\bf 124} (1989) 87.
\bibitem{nonlin1}
I.L. Buchbinder, P.M. Lavrov, {\it Classical BRST charge for nonlinear algebras},
arXiv: hep-th/0701243.
\bibitem{df} B. de Wit, D.Z. Freedman, Phys. Rev. D {\bf 21} (1980) 358.
\bibitem{tet} M. Henneaux, C. Teitelboim, {\it First and second quantized point particles of any spin},
in "Quantum mechanics of fundamental systems 2", eds. C. Teitelboim and J. Zanelli (Plenum Press, New York 1988),
p. 113.
\bibitem{fpt} A. Fotopoulos, K.L. Panigrahi, M. Tsulaia, Phys. Rev. D {\bf 74} (2006) 085029.
\bibitem{fv} E.S. Fradkin, M.A. Vasiliev, Phys. Lett. B {\bf 189} (1987) 89.
\bibitem{fv1} E.S. Fradkin, M.A. Vasiliev, Nucl. Phys. B {\bf 291} (1987) 141.
\bibitem{fads} C. Fronsdal, Phys. Rev. D {\bf 20} (1979) 848.
\bibitem{fang} J. Fang, C. Fronsdal, Phys. Rev. D {\bf 22} (1980) 1361.
\bibitem{brink} L. Brink, R.R. Metsaev, M.A. Vasiliev, Nucl. Phys. B {\bf 586} (2000) 183.
\bibitem{hull} P. de Medeiros, C. Hull, J. High Energy Phys. 0305 (2003) 019.
\bibitem{asv} K.B. Alkalaev, O.V. Shaynkman, M.A. Vasiliev, J. High Energy Phys. 0508 (2005) 069.


\end{thebibliography}
\end{document}